\documentclass[useAMS,usenatbib]{mn2e}

\usepackage{graphicx}
\usepackage{lscape}                                
\usepackage{txfonts}

\newcommand{\fastwind} {{\sc fastwind}}
\newcommand{\cmfgen} {{\sc cmfgen}}
\newcommand{\tlusty} {{\sc tlusty}}
\newcommand{\wmbasic} {{\sc wm}-{\em basic}}
\newcommand{\cloudy} {{\sc cloudy}}
\newcommand{\synspect} {{\sc synspect}}
\newcommand{\Teff} {$T_{\rm eff}$}
\newcommand{\grav} {log\,{\em g}}
\newcommand{\micro} {$\xi_{\rm t}$}
\newcommand{\kms} {km\,s$^{-1}$}
\newcommand{\nlte} {{\sc nlte}}
\newcommand{\lte} {{\sc lte}}
\newcommand{\ion}[2]{#1\,{\sc #2}}
\newcommand{\hii} {H\,{\sc ii}}
\newcommand{\Te} {$T_{\rm e}$}

\newcommand{\linean}[3]{{[\ion{#1}{#2}]\,#3}}
\newcommand{\linea}[3]{{\ion{#1}{#2}\,$\lambda$\,#3}}
\newcommand{\qion}[2]{$Q$(\ion{#1}{$^{#2}$})}
\newcommand{\xion}[3]{X(\ion{#1}{$^{#2}$}) / X(\ion{#1}{$^{#3}$})}
\newcommand{\xnion}[3]{[\ion{#1}{#2}]\,/\,[\ion{#1}{#3}]}
\newcommand{\xlinean}[6]{\linean{#1}{#2}{#3}\,/\,\linean{#4}{#5}{#6}}
\title[Ionizing radiation from massive stars and its impact on HII regions]
{The ionizing radiation from massive stars and its impact on HII regions: 
results from modern model atmospheres}
\author[S. Sim\'on-D\'iaz \& G. Stasi\'nska]{S. Sim\'on-D\'iaz$^{1,2}$
\thanks{E-mail: sergio.simon-diaz@obs.unige.ch; ssimon@iac.es} 
and G. Stasi\'nska$^{1}$\\
$^{1}$LUTH, Observatoire de Paris, CNRS, Universit\'e Paris 
           Diderot; 5 Place Jules Janssen, 92190 Meudon, France\\
$^{2}$Geneva Observatory, University of Geneva, 51 chemin 
		   des Maillettes, 1290 Sauverny, Switzerland}
\begin{document}

\date{Accepted//Received}

\pagerange{\pageref{firstpage}--\pageref{lastpage}} \pubyear{2008}

\maketitle

\label{firstpage}

\begin{abstract}
We present a detailed comparison of the ionizing spectral energy 
distributions (SEDs) predicted by four modern stellar atmosphere 
codes, \tlusty, \cmfgen, \wmbasic, and \fastwind. We consider three sets 
of stellar parameters representing a late O-type dwarf (O9.5\,V), a mid 
O-type (O7\,V) dwarf, and an early O-type dwarf (O5.5\,V). We explore
two different possibilities for such a comparison, following what we 
called \emph{evolutionary} and \emph{observational} approaches: in the 
\emph{evolutionary} approach one compares the SEDs of stars defined by the same 
values of \Teff\ and \grav; in the \emph{observational} 
approach the models to be compared do not necessarily have the same 
\Teff\ and \grav, but produce similar H and \ion{He}{i-ii} optical lines.
We find that there is a better agreement, in terms of \qion{H}{0}, 
the ratio \qion{He}{0}/\qion{H}{0}, and the shape of the SEDs predicted by 
the four codes in the spectral range between 13 and 30 eV, when models are 
compared following the \emph{observational} approach.
However, even in this case, large differences are found at higher energies. 

We then discuss how the differences in the SEDs may affect the overall 
properties of surrounding nebulae
in terms of temperature and ionization structure. We find that the effect
over the nebular temperature is not larger than 300-350 K. Contrarily,
the different SEDs produce significantly different nebular ionization 
structures. This will lead to important consequences on the establishment of
the ionization correction factors that are used in the abundance determination 
of \hii\ regions, as well as in the characterization of the ionizing stellar 
population from nebular line ratios.
\end{abstract}


%
%
\section{Introduction}
\label{section1}
%
The intense far ultraviolet (FUV) radiation emitted by early OB-type 
stars ionizes the interstellar medium, generating the so-called \hii\ 
regions. These regions can be used to derive properties of the associated 
stellar population (e.g. initial mass function, star forming rate, age), 
and other properties of the galactic region where these are located, such 
as chemical composition. However, since the properties of \hii\ regions
crucially depend on the ionizing spectral energy distribution (SED) of 
the massive star population, and this part of the stellar flux is 
generally unaccesible to direct observations, the predictions resulting 
from massive star atmosphere codes are a crucial ingredient.

The outer layers of blue luminous stars are characterized by strong 
non-LTE conditions, spherically extended geometries, and the effect of 
hundreds of thousands of flux absorbing metal lines present in the FUV  
and UV  spectral ranges (producing the so-called line-blanketing and 
line blocking effects, as well as the development of radiatively driven 
stellar winds). All the above effects must be taken into account when 
modeling the atmospheres of these stars. An ideal model atmosphere 
code should consider all of these problems in a detailed manner; however, 
this would require an enormous computational effort. To avoid this, 
depending on the specific study one wants to treat, some of the physical 
processes can be relaxed without affecting the reliability of the model 
output. 

In the last decades, a great effort has been devoted to the development 
of stellar atmosphere codes for hot massive stars. The advent of the new 
generation of non-LTE, line blanketed model atmosphere codes, either 
plane-parallel (\tlusty, Hubeny \& Lanz, \citeyear{Hub95}), or spherically
expanded (\fastwind, Santolaya-Rey et al. \citeyear{San97}, Puls et al. 
\citeyear{Pul05}; \cmfgen, 
Hillier \& Miller \citeyear{Hil98}; \wmbasic, Pauldrach et al. \citeyear{Pau01}) 
is already a fact. While \cmfgen\ aims to be the most exact one (and hence, the most
time consuming), the other three use specific approximations in the 
calculation of the stellar atmosphere structure and of the SED, reducing
then the computational time.

New generation stellar atmosphere models, including a more 
realistic description of the physical processes characterizing the 
stellar atmosphere, produce quite different ionizing SEDs than the 
previous plane-parallel, hydrostatic models (either LTE by Kurucz 
\citeyear{Kur91}, or non-LTE by e.g. Mihalas \& Auer \citeyear{Mih70}, 
Kunze et al. \citeyear{Kun92}, Kunze \citeyear{Kun94}). Some notes on this, 
and on the consequences on the 
ionization structure of \hii\ regions, can be found in Gabler et al. 
(\citeyear{Gab89}), Najarro et al. (\citeyear{Naj96}), Sellmaier et al. 
(\citeyear{Sel96}), Rubin et al. (\citeyear{Rub95}, \citeyear{Rub07}), 
Stasi\'nska \& Schaerer (\citeyear{Sta97}),  Schaerer (\citeyear{Sch00}),  
Martin-Hern\'andez et al. (\citeyear{Mart02}). Although the new predictions 
seems to go in the right direction (viz. Giveon et al. \citeyear{Giv02}, 
Morisset et al. \citeyear{Mor04}) non-negligible differences can still be 
found between the various stellar codes (see e.g. Mokiem et al. 
\citeyear{Mok04}, Martins et al. \citeyear{Mar05}, Puls et al. 
\citeyear{Pul05}).\\

In this paper, we explore in detail the differences between the
predictions in the FUV  from four modern stellar atmosphere codes
(\tlusty, \cmfgen, \wmbasic\ and \fastwind), and study their effects
on  \hii\ regions spectra. 
In Sect. \ref{section2}, we present a set of  stellar atmospheres  
constructed specifically for this purpose, and show that 
two different approaches can be followed to compare the SEDs: the 
{\em evolutionary} approach,
in which the same values of \Teff\ and \grav\ are used for models
computed with the various stellar atmosphere codes; and the 
{\em observational} approach, in which the models to be compared do 
not necessarily have the same \Teff\ and \grav, but produce similar H 
and \ion{He}{i-ii} optical lines. In Sect. \ref{section3},  by using 
simple, ab initio, 
\hii\ region models, we study how the differences in 
the SEDs may affect the overall properties of surrounding nebulae
in terms of temperature, ionization structure  and other astrophysical 
diagnostics. Finally, Sect. \ref{section4} summarizes the main 
results and presents some prospects.  

\section{Comparison of ionizing SEDs from various stellar atmosphere 
codes}\label{section2}
%
For the reader unfamiliar with stellar atmosphere modelling, the  main 
characteristics of the codes \tlusty, \cmfgen, \wmbasic\ and \fastwind\ 
are presented in the appendix (accessible on-line only).
%
\subsection{Choice of stellar atmosphere models}
\label{section21}
%
%
\begin{table}
\begin{center}
\caption{Summary of stellar atmosphere models used in this 
work (C, W, F, and T labeling \cmfgen, \wmbasic, \fastwind, and \tlusty, 
respectively).  
\label{t1}
}
\begin{tabular}{c c c c c c}    
\hline
\hline
 Models & \Teff & \grav & $R$ & log\,$\dot{M}$ & $v_{\infty}$ \\
       &  (kK) & (dex) & ($R_{\odot}$) & (M$_{\odot}$\,yr$^{-1}$) & (\kms) \\
\hline
C1, W1, F1, T1$^{\dagger}$ & 30.0 & 4.0 & \phantom{0}7.0  & -7.28 & 2000. \\
C2, W2, F2, T2$^{\dagger}$ & 35.0 & 4.0 & \phantom{0}9.0  & -7.12 & 2000. \\
C3, W3, F3, T3$^{\dagger}$ & 40.0 & 4.0 &           10.0  & -6.93 & 2400. \\
\noalign{\smallskip}
F4         & 31.0 & 4.1 & \phantom{0}7.0  & -7.28 & 2000. \\
W4         & 31.0 & 4.0 & \phantom{0}7.0  & -7.28 & 2000. \\
T4         & 30.5 & 4.1 & ---  & --- & --- \\
\noalign{\smallskip}
F5         & 36.0 & 3.9 & \phantom{0}9.0  & -7.12 & 2000. \\
W5         & 36.0 & 4.0 & \phantom{0}9.0  & -7.12 & 2000. \\
T5         & 35.5 & 4.1 & ---  & --- & --- \\
\noalign{\smallskip}
F6         & 41.0 & 4.0 &           10.0  & -6.93 & 2400. \\
\hline
\hline
\end{tabular}
\\
\end{center}
\footnotesize{$^{\dagger}$ $R$, $\dot{M}$, and $v_{\infty}$ only for \cmfgen, \wmbasic, 
and \fastwind\ models.}
\end{table}
%
Table \ref{t1} summarizes the stellar atmosphere models used for our 
study. Initially, we considered three stars with a gravity \grav\,=\,4.0 
dex and effective temperatures \Teff\,=\,30000, 35000, and 
40000 K\footnote{Following the spectral type\,-\,\Teff\ calibration by 
Martins et al. (\citeyear{Mar05}), these models would approximately represent 
a late O-type (O9.5\,V), a 
mid O-type (O7\,V), and an early O-type (O5.5\,V) dwarf, respectively.},
and computed \cmfgen, \fastwind\ and \wmbasic\ models for each pair 
\Teff\,-\,\grav. In order to exclude from our study the possibility that 
discrepancies found between the stellar atmosphere codes calculations 
could be due to differences in the considered stellar and wind parameters 
and/or stellar abundances, those models were calculated taking care of 
using exactly the same set of parameters and abundances. 

As part of the initial set of models, three \tlusty\ models, taken from 
the {\sc ostar2002} grid\footnote{See also http://nova.astro.umd.edu/} 
by Lanz \& Hubeny (\citeyear{Lan03}), were also considered (those labeled as 
G30000g400v10, G35000g400v10, and G40000g400v10, respectively). 

Finally, another set of \fastwind, \wmbasic\ and \tlusty\ models, with 
slightly modified values for the effective temperatures and gravities, 
was also computed. This was necessary for the approach presented in 
Sect. \ref{section24}.

In all the cases, the metallicity was considered to be solar
(following the set of abundances derived by Grevesse \& Sauval, 
\citeyear{Gre98}), and a microturbulence \micro\,=10\,\kms\ was 
assumed in the model calculations.
%
%
\begin{figure*}
\centering
\includegraphics[width=10 cm,angle=90]{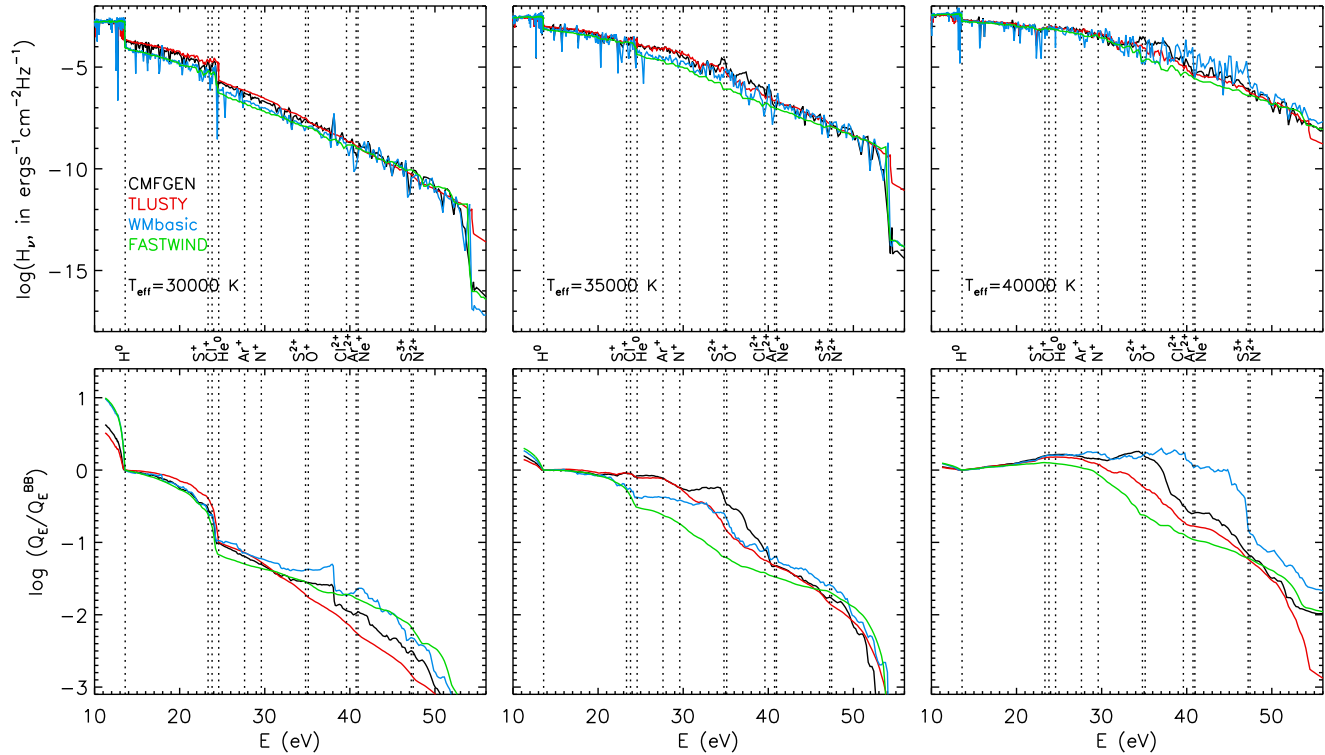}
\caption{(Upper panels) Spectral energy distribution in the range 10\,-\,60 eV as 
predicted by the various stellar atmosphere codes. \cmfgen, \tlusty, 
\wmbasic, and \fastwind\ predictions are plotted as black, red, blue, 
and green curves, respectively. The fluxes from 
\wmbasic, \cmfgen\ and \tlusty\ have been re-mapped to the \fastwind\ 
frequency grid for a better comparison. (Lower panels) Comparison of the number $Q_E$ of 
photons with energy greater than $E$ relative to the corresponding
number for a blackbody with $T$=\Teff. The location of the 
ionization edges of various ionic species of H, He, N, O, Ne, S, Cl, 
and Ar with ionization energies below 54.4 eV (\ion{He}{$^+$} ionization
energy) are also shown.
}
\label{fig1}
\end{figure*}
%
%
\subsection{Methodology}
\label{section22}
%
Two different approaches were followed to compare the SEDs from the 
various stellar atmosphere codes. Basically, these are characterized by
the methodology used to establish the stellar parameters (\Teff, \grav, 
and log\,$L$). We call them {\em evolutionary} and 
{\em observational} approaches. 

In the {\em evolutionary approach} the stellar parameters characterizing a
star are fixed by the output of stellar evolutionary codes. This approach 
is mainly used by stellar population synthesis codes, such as 
{\sc starburst99} (Leitherer 
\citeyear{Lei99}, see also Smith et al. \citeyear{Smi02}) or {\sc pegase} 
(Fioc \& Rocca-Volmerange \citeyear{Fio99}). In this case, for a given 
initial stellar mass and evolutionary time, a star is defined by the 
\Teff, \grav, and log\,$L$ predicted by the stellar evolution codes 
(e.g. Schaller et al. \citeyear{Sch92}; Meynet \& Maeder \citeyear{Mey00}, 
\citeyear{Mey05}).

On the other hand, when spectroscopic and photometric observations of 
the star are available, the stellar parameters can be derived by means 
of what we call an {\em observational approach}. In this case, the 
stellar and wind parameters are determined by fitting appropriate 
stellar line features using stellar 
atmosphere models. This is a long established way to analyze O-type 
star spectra. The optical H and \ion{He}{i-ii} 
lines have been mainly used to determine the stellar \Teff\ and \grav\ 
(viz. Herrero et al. 
\citeyear{Her92}, \citeyear{Her02}; Repolust et al. 2004), though the UV 
spectra of these stars also provide valuable information about the 
physical properties of their winds, such as terminal wind velocities 
and mass loss rates (see Kudritzki \& Puls \citeyear{Kud00}).
In addition, in the observational approach, the stellar luminosity is 
obtained from the stellar absolute visual magnitude ($M_{\rm v}$).\\

For a quantitative comparison of ionizing SEDs, we used the quantity 
$Q_E$, indicating the number of photons with energy greater than E=h$\nu$. 
If the emergent stellar flux is given by $H_{\nu}$($\lambda$), these 
quantity is defined as follows:
%
\begin{equation}\label{Eq1}
Q_E\,=\,\int_{\nu}^{\infty}\,\frac{L_{\nu}}{{\rm h}\nu'}\,d\nu'\,=\,16\pi^2\,R^2\,
\int_{0}^{\lambda}\,\frac{H_{\nu}(\lambda')}{{\rm h}\lambda'}\,{\rm d}\lambda'
\end{equation}
%
where $L_{\nu}$ the emergent stellar luminosity distribution, and $R$ the 
stellar radius.
%
\subsection{Comparison of SEDs in the evolutionary approach}
\label{section23}
%
In this section, we compare the SEDs resulting from \cmfgen, \tlusty, 
\wmbasic, and \fastwind\ when models with exactly the same stellar parameters are 
considered (\Teff, \grav\ and $L$). Through this comparison we can test 
the implications of using SEDs from the various stellar atmosphere codes 
as input for stellar population synthesis codes in order to compute 
photoionization models of giant \hii\ regions.\\

Upper panels in Figure \ref{fig1} show the corresponding SEDs in the range
10\,-\,60\,eV. This qualitative comparison already shows that 
the stellar atmosphere codes produce somewhat different 
ionizing SEDs. To quantify these differences, we first consider the total
number of \ion{H}{$^0$} ionizing photons, and then compare the shape of the 
SEDs for energies $E\,>\,$13.6 eV.

%
\subsubsection{Ionizing luminosities}
\label{section231}
%
%
\begin{table}
\begin{center}
\caption{Resulting number of H$^0$ ionizing photons (in
logarithm) for the various stellar atmosphere codes when the same 
\Teff\ and \grav\ are considered (\emph{evolutionary} approach).
\label{t2}
}
\begin{tabular}{c | c  c  c}    
\hline
\hline
 \Teff   &  30000  &  35000  & 40000  \\
 \grav   & 4.0  &  4.0 &  4.0 \\
 log\,$L$   & 38.10  &  38.55 &  39.80 \\
\hline
C1, C2, C3 & 47.33 &  48.39 &  48.99 \\
T1, T2, T3 & 47.43 &  48.44 &  49.00 \\
W1, W2, W3 & 46.97 &  48.30 &  48.93 \\
F1, F2, F3 & 46.98 &  48.26 &  48.92 \\
\hline
\hline
\end{tabular}
\\
\end{center}
\end{table}
%
Table \ref{t2} summarizes the number of \ion{H}{$^0$} ionizing photons,
\qion{H}{0}\,=\,$Q_{13.6}$, 
resulting from the various stellar atmosphere models. 
Note that in this case, the stellar radii (used in the calculation of the
ionizing luminosities by means of Eq. \ref{Eq1}) were fixed 
once a characteristic stellar luminosity was assumed for each effective 
temperature. 

The values of log\,\qion{H}{0}\ computed with the various codes for the same  
\Teff, \grav, log\,$L$ are found to differ significantly. In the worst case, 
i.e. between \wmbasic\ 
and \tlusty\ for \Teff=30000, they differ by  $\sim$ 0.5 dex. The discrepancy 
tends to be lower for larger \Teff, but  in the hottest star a difference 
of 0.08 dex (i.e. a factor of 1.2) can still be found between \fastwind\ and 
\cmfgen\ models. Note, however, that there is always a good agreement 
(below 0.04 dex) between \wmbasic\ and \fastwind\ models on one hand, and 
\cmfgen\ and \tlusty\ models on the other. 
%
\subsubsection{Shape of the ionizing SEDs}
\label{section232}
%
To study how the shape of the predicted SEDs compares in the H Lyman 
continuum\footnote{Note that shocks were not considered in any of the 
models and consequently the emergent stellar flux in the X-ray 
domain is virtually zero.} we produced the diagrams shown in lower 
panels of Figure \ref{fig1}. There we compare, for each energy $E$ (eV), 
the number $Q_E$ of photons with energy greater than $E$ relative to 
the corresponding number for a blackbody with $T$=\Teff.
Note that all spectra were scaled to have the same value for $Q_{13.6}$.

These diagrams can also be used to compare the number of ionizing photons 
for a certain ionic species, \ion{X}{$^i$}, relative to the total number 
of hydrogen ionizing photons -- \qion{X}{i}/\qion{H}{0} --. This is the major factor 
determining the relative populations of different ions of a given element 
in an \hii\ region, although its effect is modulated by the recombination and 
charge exchange coefficients. 

Since we are interested on the impact that the ionizing radiation from 
massive stars has on \hii\ regions, we only considered ions of those 
elements which lines are commonly observed in \hii\ regions (i.e.
H, He, N, O, Ne, S, Cl, Ar). The corresponding ionization potentials
are indicated as vertical lines in Figure \ref{fig1}.
Note that, for the range of effective temperatures we are considering, all the SEDs 
show an abrupt decrease at energies $>$54.4\,eV. The effect of this part 
of the stellar flux on the ionization of the surrounding material in
the case of \hii\ regions is practically negligible, so we restrict our 
study to ions with ionization potential $<$54.4\,eV. \\

Various conclusions concerning the shape of the SEDs can be emphasized 
from the inspection of Figure \ref{fig1}:
\begin{description}
\item [{\em 30000 K star:}]{In this case \fastwind\ and 
\wmbasic\ models predict a larger H Lyman jump than the other 
two codes. This is consistent with the fact
that a lower number of \ion{H}{$^0$} ionizing photons is resulting
from \fastwind\ and \wmbasic\ models (see above).

Good agreement is found for all the codes
except \tlusty\ in the region between the \ion{H}{$^0$} and 
\ion{He}{$^0$} edges. This code is producing a somewhat harder 
flux, and hence larger \qion{S}{+} and \qion{Cl}{+} values.

While \cmfgen\ and \wmbasic\ predict a similar \ion{He}{$^0$} jump,
this break is larger in \tlusty\ and \fastwind\ SEDs. As a consequence,
\tlusty\ results in the same \qion{He}{0} than \fastwind\
and \wmbasic, and \fastwind\ in a smaller number of \ion{He}{$^0$}
ionizing photons ($\sim$\,0.2 dex).

The slope of the SED above the \ion{He}{$^0$} edge is larger in \tlusty\ and
\cmfgen\ than in \wmbasic\ and \fastwind, which results in harder
fluxes for these later codes for energies $E\,>\,$40 eV.

It is also interesting to note how the presence of a strong emission 
line in $\sim$\,38 eV (present in \wmbasic\ and \cmfgen\ spectra, but not in 
\fastwind\ and \tlusty) affects the resulting \qion{S}{2+} and 
\qion{O}{+}.\\}

\item [{\em 35000 K star:}]{The agreement between \tlusty\ and \cmfgen\ is 
quite good, with discrepancies not larger than 0.1 dex in $Q_E$, except 
in the range 30\,-\,40\,eV. This is a region of the stellar flux 
where metal line blocking is important, and differences in the way this 
is treated (e.g. the amount of metal lines included in the blocking 
calculation), or differences in the ionization degree of the elements 
producing this forest of lines can affect the amount of flux emerging 
from the star in this spectral range.
The discrepancy found between \cmfgen\ and \tlusty\ in this spectral
range translates in \qion{S}{2+} and \qion{O}{+} values differing 
by $\sim$\,0.4 dex.

Important discrepancies are found between these two codes and 
\wmbasic\ and \fastwind\ in the spectral range where the \ion{S}{$^+$}, 
\ion{Cl}{$^+$}, \ion{He}{$^0$}, \ion{Ar}{$^+$}, \ion{N}{$^+$} ionizing edges
are located ($\sim$\,23\,-\,30 eV). Both \wmbasic\ and \fastwind\ result 
in lower fluxes in this spectral range. For example, \qion{He}{0}/\qion{H}{0}
is $\sim$\,0.3\,-\,0.4 dex smaller in \fastwind\ and \wmbasic\ than in the
other codes. In addition, while \wmbasic\ is closer
to \cmfgen\ and \tlusty\ for larger energies, \fastwind\ is producing 
a lower flux between 30 and 45 eV. 

In this case \wmbasic\ and \fastwind\ are also resulting
is a somewhat larger H Lyman edge (again consistent with results in
Table \ref{t2}).\\}

\item [{\em 40000 K star:}]{Good agreement is found between \cmfgen\
and \tlusty\ (again except in the range 30\,-\,40\,eV). \fastwind\
predicts a somewhat lower flux than these two codes for E\,$\ge$\,25\,eV. 
On the other hand, while \wmbasic\ results in a similar SED as \cmfgen\ for 
E\,$\le$\,35\,eV, differences of up to 0.8 dex in $Q_E$ are found for larger 
energies.}

\end{description} 
%
\subsection{Comparison of SEDs in the observational approach}
\label{section24}
%
In this section we check if the various stellar atmosphere codes are
producing similar results for the lines normally used to estimate
the stellar parameters. Then we compare the SEDs given by models 
which do not have necessarily the same \Teff\ and \grav\ but produce 
similar results for those lines.
%
\subsubsection{Predicted stellar HHe optical lines}
\label{section241}
%
%
\begin{figure*}
\centering
\includegraphics[width=8.0 cm,angle=90]{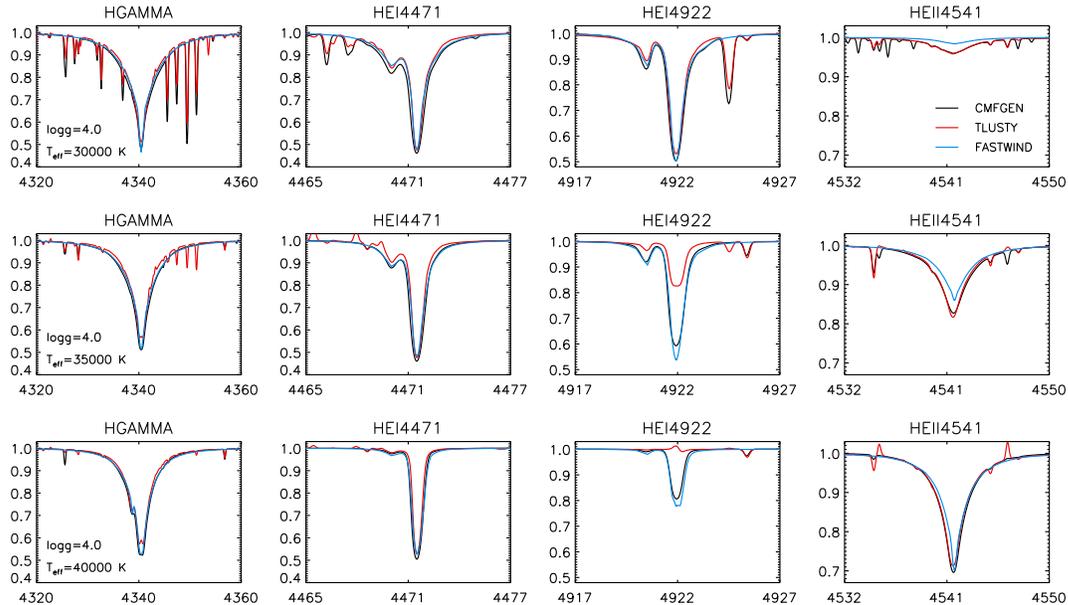}
\caption{Comparison of some of the optical H, \ion{He}{i}, and
\ion{He}{ii} lines commonly used to derive stellar parameters
in O-type stars resulting from \cmfgen, \tlusty, and \fastwind\
models. Note that only H and He lines are present in the 
\fastwind\ synthetic spectra (see appendix).}
\label{fig2}
\end{figure*}
%
Figure \ref{fig2} shows the comparison of a representative set 
of H Balmer and \ion{He}{i-ii} synthetic lines for the various 
values of  (\Teff, \grav), as predicted by \fastwind, 
\cmfgen, and \tlusty. We decided to present results in two \ion{He}{i} 
lines to account for the singlet-triplet problem (see Najarro et 
al. \citeyear{Naj06}, and references therein).
\wmbasic\ is not included in this comparison since the optical 
H and \ion{He}{i-ii} lines are not synthesized by this code. 

%
%
\begin{table}
\begin{center}
\caption{Stellar parameters (\Teff\ and \grav) required to 
fit \fastwind\ and \tlusty\ optical H and He lines to those resulting from \cmfgen\ 
models with \grav\,=4.0 and 
\Teff\,=\,30000, 35000, and 40000 K, respectively. The labels indicated 
in brackets are the same as those used in Table \ref{t1}  
\label{t3}
}
\begin{tabular}{c c c}    
\hline
\hline
 \cmfgen & \fastwind & \tlusty \\
\hline
30000, 4.0 (C1) & 31000, 4.1 (F4) & 30500, 4.1 (T4) \\
35000, 4.0 (C2) & 36000, 3.9 (F5) & 35500, 4.1 (T5) \\
40000, 4.0 (C3) & 41000, 4.0 (F6) & 40000, 4.0 (T3) \\
\hline
\noalign{\smallskip}
\end{tabular}
\\
\end{center}
\end{table}
%

While \fastwind\ and \cmfgen\ produce quite similar 
H$_{\gamma}$ wings (which are the main diagnostic to determine  \grav), 
\tlusty\ models would need a larger \grav\ to produce the same result.
On the other hand, while the three 
codes predict coherent results for the \linea{He}{i}{4471} (triplet)
line, this is not the case for the other \linea{He}{i}{4922} (singlet) 
line. The cause of this discrepancy was already pointed out by Najarro 
et al. (\citeyear{Naj06}), who also suggested that the
\ion{He}{i} triplet lines should be preferred in the spectral analysis
due to the sensitivity of the \ion{He}{i} singlet transitions to model
assumptions.
Finally, \cmfgen\ and \tlusty\ agree almost perfectly 
regarding the \linea{He}{ii}{4541} line, while \fastwind\ models 
result in a shallower line in the \Teff\,=\,30000 and 35000 K 
cases. Although it is not shown here, a similar behavior is also found 
for the other \ion{He}{ii} lines.

The above-mentioned result implies\footnote{Villamariz \& Herrero 
(\citeyear{Vil00}) showed that, except for \linea{He}{ii}{4686}, the 
\ion{He}{ii} lines are practically insensitive to the adopted value 
of microturbulence. Note also that for effective temperatures in 
the range between 30000 and 35000 K, the sensitivity of the 
\ion{He}{i} lines on \Teff\ is quite small.} that the codes are 
resulting in a somewhat different temperature and density structure
in the stellar atmosphere, which in turn affect the strength of the
\ion{He}{ii} lines (via the \ion{He}{} ionization degree in the line 
formation region) and the wings of the H Balmer lines. On the
other hand, the stellar parameters that would be derived from 
the spectroscopic analysis of a given observed spectrum are 
model dependent. To explore the differences
in the stellar parameters that would be obtained when using the various
codes we obtained the \Teff\ and \grav\ values required by \fastwind\ 
and \tlusty\ to produce similar optical H and \ion{He}{i-ii} than \cmfgen\ 
models, taken as reference. These are indicated in Table \ref{t3}.  

The discrepancies between the three codes for the 30000\,K and 35000\,K 
stars are $\sim$\,500\,-\,1000 K in \Teff\ and 0.1 dex in \grav. Note
that these discrepancies are of the same order as the accuracy reached 
when analyzing \emph {observed} spectra of OB-type stars with a given 
code ($\sim$\,1000\,K in \Teff\ and 0.1 dex in \grav\ --- see e.g. 
Herrero et al. \citeyear{Her02}, Repolust et al. \citeyear{Rep04}, 
Sim\'on-D\'iaz et al. \citeyear{Sim06}). On the other hand, the agreement
for the 40000 K star case is almost perfect.

In next sections, we take into account these differences for the comparison 
of ionizing SEDs.
%
\subsubsection{Ionizing luminosities}
\label{section242}
%
%
%
\begin{table}
\begin{center}
\caption{Number of H$^0$ ionizing photons for \cmfgen, \tlusty, \fastwind, 
and \wmbasic\ models with different stellar parameters (\Teff\ and \grav), 
but similar H and He optical lines. The stellar radius was determined 
following Kudritzki et al. (\citeyear{Kud80}), once an absolute visual 
magnitude ($M_{\rm v}$) was assumed for each set of models.
\label{t4}
}
\begin{tabular}{c | c | c | c c c c}    
\hline
\hline
  & \multicolumn{6}{c}{$M_{\rm v}$\,=\,-3.5} \\
\cline{2-7}
\noalign{\smallskip}
  & \Teff, \grav   & V      & $R$/$R_{\odot}$ & log\,$L$ & $M$/$M_{\odot}$ & 
log\,\qion{H}{0}\ \\
\hline
C1 & 30000, 4.0 & -29.09 & 6.3 & 38.05 & 14.3 & 47.28 \\
T4 & 30500, 4.1 & -29.09 & 6.2 & 38.07 & 17.8 & 47.45 \\
F4 & 31000, 4.1 & -29.05 & 6.3 & 38.11 & 18.5 & 47.40 \\
\hline
\noalign{\smallskip}
\hline
  & \multicolumn{6}{c}{$M_{\rm v}$\,=\,-4.3} \\
\cline{2-7}
\noalign{\smallskip}
 &  \Teff, \grav & V     & $R$/$R_{\odot}$ & log\,$L$ & $M$/$M_{\odot}$ & 
log\,\qion{H}{0}\ \\
\hline
C2 & 35000, 4.0 & -29.37 & 7.9 & 38.52 & 22.8 & 48.36 \\
T5 & 35500, 4.1 & -29.37 & 7.9 & 38.54 & 28.8 & 48.45 \\
F5 & 36000, 3.9 & -29.37 & 7.9 & 38.57 & 18.2 & 48.40 \\
\hline
\noalign{\smallskip}
\hline
  & \multicolumn{6}{c}{$M_{\rm v}$\,=\,-4.8} \\
\cline{2-7}
\noalign{\smallskip}
  &  \Teff, \grav & V & $R$/$R_{\odot}$ & log\,$L$ & $M$/$M_{\odot}$ & log\,\qion{H}{0}\ 
\\
\hline
C3 & 40000, 4.0 & -29.55  & 9.2 & 38.88 & 31.0  & 48.97 \\
T3 & 40000, 4.0 & -29.52  & 9.3 & 38.89 & 31.8  & 49.00 \\
F6 & 41000, 4.0 & -29.56  & 9.2 & 38.92 & 30.6  & 48.99 \\
\hline
\hline
\end{tabular}
\\
\end{center}
\end{table}
%
Table \ref{t4} indicates the number of \ion{H}{$^0$} ionizing 
photons for the models summarized in Table \ref{t3}. The  synthetic V 
magnitudes resulting from the models, along with
the derived stellar radii, luminosities and masses are also
indicated. Note that in this case,
the stellar radii were determined following Kudritzki et al. (\citeyear{Kud80}), 
once an absolute visual magnitude ($M_{\rm v}$) was assumed for each set of models.

In all cases, the agreement between models which results in similar H and 
\ion{He}{i-ii} lines is better than when the same stellar parameters are 
used to construct the models (see Table \ref{t2}). 
Note, however, that differences up to 0.1 dex still remain. These
differences are partially due to the somewhat different stellar
radii (which are a consequence of the synthetic V magnitude resulting
from the stellar models).
%
\subsubsection{Shape of the ionizing SEDs}
\label{section243}
%
%
%
\begin{figure*}
\centering
\includegraphics[width=5.5 cm,angle=90]{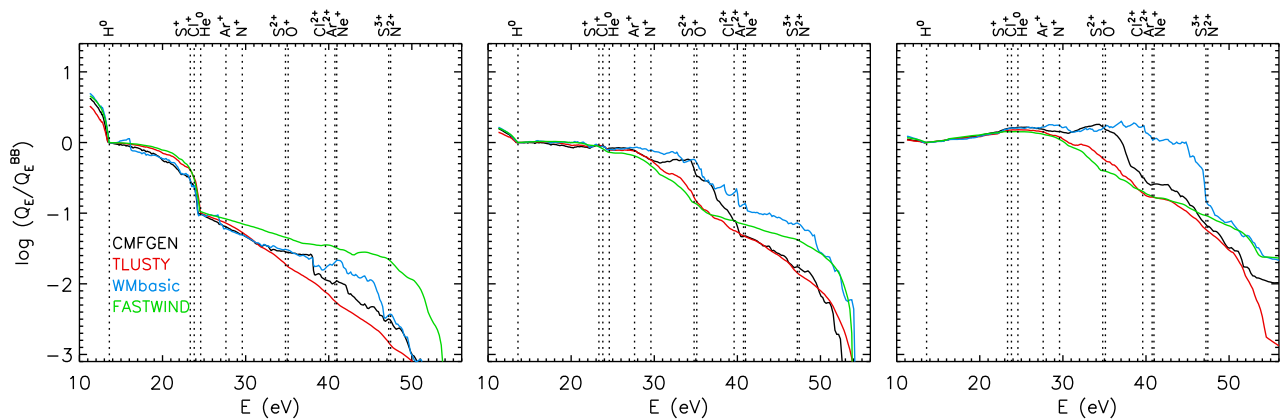}
\caption{As lower panels of Fig. \ref{fig1} but for the \cmfgen, \tlusty, and
\fastwind\ models resulting in similar optical H and \ion{He}{i-ii} lines
(see Sect. \ref{section24}). The \wmbasic\ models appearing in these plots
were selected to produce similar SEDs as the other codes in the range
13.6\,-\,30 eV (see explanation in Sect. \ref{section25}).}
\label{fig3}
\end{figure*}
%
Figure \ref{fig3} compares the shape of the SEDs for those \cmfgen, \tlusty\ 
and \fastwind\ models resulting in similar H and \ion{He}{i-ii} lines. 


In all the cases, the discrepancies found in the spectral range 13.6\,-\,30 eV 
when models with the same stellar parameters were compared (see Sect. 
\ref{section23} and Fig. \ref{fig1}) have practically disappeared. 
This is especially noticeable in the 35 and 40 kK star cases. In addition, the 
agreement between the three codes for the H Lyman break is quite good.

Although \wmbasic\ could not be included in the comparison presented in 
Section \ref{section241} (since the optical H and \ion{He}{i-ii} lines 
are not sythesized by this code), in view of the abovementioned results,
we decided to also include the SED of those \wmbasic\ models which, having
an effective temperature and gravity close to those of \cmfgen\ models,
produce similar SEDs as the other codes in the range
13.6\,-\,30 eV. These models correspond to those labeled W4, W5, 
and W3 in Table \ref{t1}.\\
\newline
In general, the situation for higher energies has also improved; however
some important discrepancies still remain. From inspection of Fig. 
\ref{fig3} it can be remarked that:
\begin{itemize}
\item {\fastwind\ gives much harder SEDS than the other codes in the
30000 K  case}
\item {\wmbasic\ results in harder fluxes than the other codes
in the 35000 and 40000 K cases (this discrepancy is
especially noticeable in the hottest star).}
\item {\tlusty\ and \cmfgen\ SEDs agree quite well, except in the range
30\,-\,40\,eV. In this energy range, the \fastwind\ SED is closer to that
of \tlusty.}
\end{itemize} 
As commented by Puls et al. (\citeyear{Pul05}), the number of metal lines
between 35 and 50\,eV is quite large, 
and the treatment of the opacities of the weakest lines affects the SED 
above 24.6 eV. The different approaches of the line blanketing and wind 
effects might also contribute to the observed discrepancies. In addition,
the fact that different effective temperatures are being considered
in the various codes may affect the slope of the SED in the high energy
range.
%
\subsection{The importance of the HHe line analysis in the comparison of
SEDs}\label{section25}
In the previous section we have shown that the number of \ion{H}{$^0$} 
ionizing photons, the ratio \qion{He}{0}/\qion{H}{0}, and the
shape of the SEDs below 30\,eV agree better in 
the observational approach than in the evolutionary approach. 
This results from the fact that the shape of the 
emergent flux is a consequence of the stellar atmospheric structure 
(which, for a given set of stellar parameters, is influenced by the 
line blanketing and wind-blanketing effects) and the blocking by 
numerous metal lines present in the FUV  spectral range. 
Specifically, the size of \ion{H}{$^0$} and \ion{He}{$^0$} jumps are a 
function of the population of the ground levels of the corresponding 
ions in the respective continuum forming regions.
Temperature and electron density mainly control the level population of those 
ground levels (though wind effects can also contribute, see Gabler 
et al \citeyear{Gab89}, and Najarro et al. \citeyear{Naj96}). 
For example, in a star with larger effective temperature or lower
gravity, H and He are expected to be more ionized in the corresponding 
continuum forming regions, and hence the ground level population 
of \ion{H}{$^0$} and \ion{He}{$^0$} will be smaller, resulting in 
smaller H and \ion{He}{i} jumps.

On the other hand, the optical H and \ion{He}{i-ii} lines depend mainly on the
recombination from the corresponding upper ions (i.e. \ion{H}{$^+$},
\ion{He}{$^+$} and \ion{He}{$^{2+}$}, respectively) which, in
turn, is controlled by temperature and electron density, again in the 
region of the stellar atmosphere where these transitions become
optically thin. These line formation regions are normally located 
close to the regions where the \ion{H}{$^0$} and \ion{He}{$^0$} 
continua are formed, except maybe for H$_{\alpha}$, 
\linea{He}{ii}{4686}, \linea{He}{i}{5875}, the strongest lines.

By fitting the optical H and \ion{He}{i-ii} lines, one is fixing
the physical properties (density and temperature) in the corresponding 
line forming region, and hence in the \ion{H}{$^0$} and \ion{He}{$^0$} 
continuum forming regions. 
Consequently, differences in the predicted SEDs, in terms of 
\ion{H}{$^0$} Lyman jump and \qion{He}{0}/\qion{H}{0} ratio, and
differences in the optical H and \ion{He}{ii} lines between any
couple of models are strongly related.

Therefore, the study of the optical H and \ion{He}{i-ii} stellar lines 
provide very useful additional information to understand possible 
discrepancies in the ionizing stellar SEDs predicted by the various 
stellar atmosphere codes. 
%
\subsection{Use of SpT-\Teff\ calibrations}
\label{section26}
%
In section \ref{section22} we have presented two different approaches
to establish the stellar parameters characterizing a star. There is still
another way, based on SpT-\Teff\ calibrations.

It is important to note that while the spectral classification of a star 
provides a qualitative description of the stellar characteristics, and 
does not depend of any model calculation, the stellar parameters 
determination is dependent on the physics used in the stellar atmosphere 
modeling. For example, in the last years there has been an important revision 
of the calibrations of stellar parameters of O-type stars (see e.g. Martins 
et al. \citeyear{Mar05}). 
The inclusion of non-LTE, line blanketing and wind effects in the stellar 
atmosphere code calculations resulted in SpT\,-\,\Teff\ calibrations 
indicating a lower effective temperature for a given spectral type compared to 
previous calibrations (Vacca et al. \citeyear{Vac96}).

The above results (Sects. \ref{section24}) alerts us that one 
must use the SpT-\Teff\ calibrations with care, because these are 
model dependent. Obviously, this does not mean that  
calibrations cannot be used at all, but it is important to understand
that applying a calibration obtained with one code to create a grid of 
stellar SEDs for O-type stars with a different code would be an inconsistent way of 
doing.
Moreover, even using the same code, it should be reminded that a certain
dispersion is always associated with a given calibration.
We refer the reader to Martins et al. (\citeyear{Mar05}) and Mokiem et 
al. (\citeyear{Mok04}) for a detailed discussion on the dependence of the 
SpT\,-\,\Teff\ calibrations on model assumptions such as microturbulence,
metallicity, and atmospheric and wind parameters. As commented by
Martins et al. (\citeyear{Mar05}), their theoretical \Teff\ scale should
be taken as indicative since it is expected to have an uncertainty of 
$\pm$1000 to 2000 K due to both a natural dispersion and uncertainties 
inherent to the methodology they applied. 
%
\section{Predicted effects of different SEDs on surrounding nebulae}
\label{section3}
%
In this section, we investigate how the differences in the SEDs may 
affect the overall properties of surrounding ionized nebulae. For illustrative 
purposes, we have constructed  a set of ab initio photoionization models
with \cloudy\footnote{We used version 07.02 of Cloudy, last described by 
Ferland et al. (\citeyear{Fer98}).}, using as an input the SEDs from the 
stellar atmosphere models discussed above. 

To isolate the dependence of the nebular ionization structure on the stellar 
SED from the other factors that may also affect it, such as nebular gas 
distribution and ionization parameter, we limited ourselves to spherical constant 
density models ($N_{\rm H}$\,=\,1000 cm$^{-3}$, $R_{\rm int}$\,=\,10$^{15}$ cm), 
and the same \qion{H}{0} was 
considered for each set of stellar parameters. The adopted nebular abundances 
were those of the Orion nebula and all the models are dust-free.
%
\subsection{Effect on the nebular temperature}
\label{section31}
%
\subsubsection{General considerations}
\label{section311}
%
At each pointin the nebula, the electron temperature, \Te,  is the result of 
balance between heating and cooling 
processes. The energy gains in a photoionized nebula are usually dominated by 
photoionization of hydrogen atoms, with some contribution of photoionization 
of helium (Osterbrock 1989; see also review by Stasi\'nska \citeyear{Sta04}). 
It can be shown that, when ionization equilibrium is achieved, the energy gains 
due to ionization of \ion{H}{$^0$} and \ion{He}{$^0$} can, in a first 
approximation,  be written as:
\begin{equation}\label{equation2}
G = n_{\rm H}^{+}n_e\ (<E_{\rm H}>\alpha_{\rm H} 
                     + <E_{\rm He}>\alpha_{\rm He} 
					 \frac{n_{\rm He}}{n_{\rm H}}
                     \frac{n_{\rm He}^+}{n_{\rm He}}
					 \frac{n_{\rm H}}{n_{\rm H}^+}),
\end{equation}
where $<E_{\rm H}>$ is the average energy gained per photoionization
of an \ion{H}{$^0$} atom:
\begin{equation}\label{equation3}
<E_{\rm H}> = \frac{\int^{\infty}_{h{\nu}_0} 4\pi \frac{J_{\nu}(r)}{h\nu}
                           e^{-\tau_{\nu}(r)}a_{\nu}({\rm H}^{0})(h\nu - h\nu _{0})dh\nu}
						   {{\int^\infty_{h\nu_0}4\pi \frac{J_{\nu}(r)}{h\nu}
                           e^{-\tau_{\nu}(r)}a_{\nu}(\ion{H}{^0})dh\nu}},
\end{equation}
with $J_{\nu}$ being the intensity of the ionizing radiation.
An  expression similar to Eq. \ref{equation3} can be written for $<E_{\rm He}>$. \\

Therefore, $<E>$ depends on the shape of the ionizing SED, but not on the 
stellar luminosity, nor on distance to the ionizing source.\\

On the other hand, the most important source of cooling in the nebula is generally 
collisional excitation of low-lying energy levels of abundant elements
(O, S, N) followed by radiative de-excitation.  
%
\subsubsection{Quantitative effects of different SEDs on \Te}
\label{section312}
%
%
%
\begin{table}
\begin{center}
\caption{Nebular temperature averages (over volume) weighted 
by the electron density predicted by \cloudy\ models using the SEDs from 
the stellar models considered in Sect. \ref{section23}  
\label{t5}
}
\begin{tabular}{c c c c c c c}    
\hline
\hline
 &   \Te & \Te(\ion{O}{$^{2+}$}) & \Te(\ion{O}{$^+$})  \\
\hline
C1 & 7100 & 6960 & 7100  \\
T1 & 7340 & 7240 & 7340 \\
W1 & 7060 & 6890 & 7070 \\
F1 & 7040 & 6910 & 7040 \\
\noalign{\smallskip}
C2 & 7580 & 7410 & 7760 \\
T2 & 7820 & 7670 & 7930 \\
W2 & 7630 & 7480 & 7730 \\
F2 & 7750 & 7630 & 7790 \\
\noalign{\smallskip}
C3 & 8280 & 8170 & 9100 \\
T3 & 8340 & 8200 & 8880 \\
W3 & 8190 & 8070 & 8960 \\
F3 & 8360 & 8170 & 8690 \\
\hline
\hline
\end{tabular}
\\
\end{center}
\end{table}
%
Table \ref{t5} summarizes various nebular temperature averages (over volume and
weighted by the electron density) predicted by
\cloudy\ models using the SEDs from the stellar models considered in Sect.
\ref{section23}.

The first thing that can be noted is that, in spite of the large differences
found between some of the SEDs (see Fig. \ref{fig1}), the resulting nebular electron
temperatures do not differ by more than 300\,-\,350 K. 

\begin{figure}
\centering
\includegraphics[width=8.0 cm,angle=0]{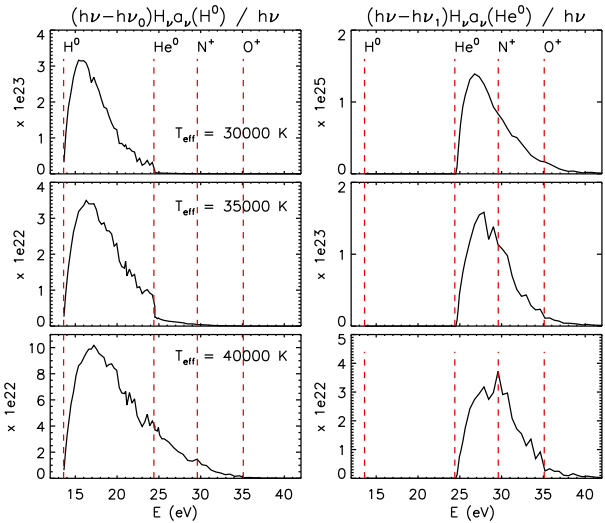}
\caption{Numerator functions to be integrated to calculate $<E_{\rm H}>$ 
         (left) and $<E_{\rm He}>$(right) from Eq. \ref{equation3}. 
		 Note the different scales used in each plot.}
\label{fig4}
\end{figure}

To understand this one should
note that in the calculation of $<E>$ from Eq. \ref{equation3}, the shape of the 
stellar SED is modified by the 
effect of absorption of stellar radiation in intervening nebular layers, 
e$^{-\tau_{\nu}}$, and by the absorption cross section, $a_{\nu}$(\ion{H}{$^0$}). 
Therefore, the contribution of different spectral regions from the SED to the 
nebular heating is not equally important.
The functions to be integrated in the numerator of Eq. \ref{equation3}, for the case of 
an
optical thickness of zero\footnote {Inside the nebula, those functions are affected by 
absorption, and by the contribution of the diffuse ionizing radiation produced in the 
nebula, both of which are dependent on the SED only to second order.}, are plotted in 
Fig. \ref{fig4}, with the left 
panels corresponding to H and the right panels corresponding to He. 
\fastwind\ stellar SEDs for stars with \Teff\,=\,30000, 35000 and 40000 K 
(from top to bottom, respectively) were used as illustrative. 
These figures show that the average energy gained per photoionization 
of \ion{H}{$^0$} in the nebula is independent of the SED above 25 eV 
for the stars with \Teff\,=\, 30000 and 35000 K, and 35 eV  for the 
\Teff\,=\,40000 K case. Consequently, if 
the SEDs predicted by the stellar atmosphere codes differ below these 
energy limits, the nebular electronic temperature may be different. On the 
other hand, differences in the SEDs above these limits will not have any 
effect on the heating of the nebular gas. A similar argument can be applied to 
the heating by photoionization of \ion{He}{$^0$}, which is dominated by 
photons with energies around 30 eV. Note, however, that the latter
is always much smaller than  \ion{H}{$^0$} heating, 
because of the $n_{\rm He}/n_{\rm H}$ factor in Eq. \ref{equation2}. 

%

One could then expect that, for a photoionization model
with the same nebular abundances, the harder the ionizing fluxes for energies
between 15-30 eV, the higher the nebular electron temperature. However,
from inspection of the first column of Table \ref{t5}, it can be seen that
actually this is not the general behavior. For example, in the \Teff\,=\,35000 K case,
while \cmfgen\ is resulting in the hardest ionizing flux (see Fig. \ref{fig1}),
the corresponding photoionization model is the one giving the lowest \Te.
On the other hand, for the same star case, in spite of the fact that 
\cmfgen\ and \tlusty\ are producing a similar heating, there is a difference 
in \Te\ between the associated \cloudy\ models of 240 K.\\ 
 
The reason for this is that, although the nebular abundances are the same,
the various SEDs are producing a different nebular ionization structure
and hence the ionic fractions of \ion{O}{$^{2+}$}, \ion{O}{$^+$}, 
\ion{S}{$^{3+}$}, \ion{S}{$^{2+}$}, \ion{S}{$^{+}$}, and \ion{N}{$^+$}
(main coolants) are different. For example, in the first case mentioned
above, since the number of \ion{O}{$^{+}$} ionizing photons is larger
in \cmfgen\ (see Fig. \ref{fig1}), the population of \ion{O}{$^{++}$} will be 
larger in the nebula. Since \linean{O}{iii}{} optical and infrared lines 
are very efficient coolants, the overall cooling will be more effective. 
Note that this effect is so
strong that the resulting nebular \Te\ in the \ion{O}{$^{2+}$} region
(and hence in the global nebula) is lower than for the nebula ionized 
by \fastwind\ model, which has a weaker ionizing flux, but the amount 
of \ion{O}{$^{+}$} ionizing photons is smaller.

Because of this effect, the shape of the SED in the spectral range containing the 
\ion{O}{$^{+}$} and \ion{S}{$^{2+}$} edges is crucial. This range
is precisely where the larger dispersion between the various SEDs
is found, even when the fluxes are compared in the \emph{observational} 
approach (see Fig. \ref{fig3}).

Obviously, the magnitude of these effects depend on the star and on the 
nebula under consideration. For example, in the
\Teff\,=\,30000 K case the \ion{O}{$^{2+}$} region is very small, and
hence the global effect of cooling by \linean{O}{iii}{} lines is not very
important. In this case, both \Te(\ion{O}{$^{+}$}) and \Te\ are similar,
and the nebula with a higher \Te\ is the one ionized by \tlusty, whose
stellar flux is the hardest in the 20-25 eV region.\\

%
%
\begin{figure}
\centering
\includegraphics[width=9.5 cm,angle=90]{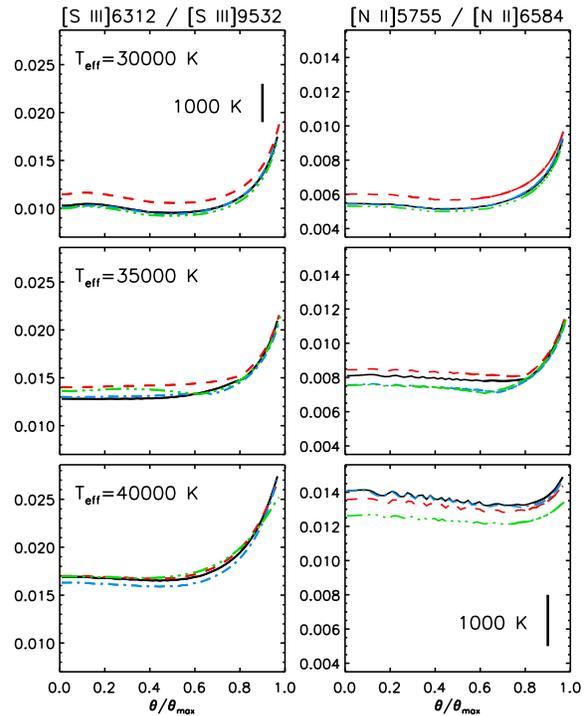}
\caption{Nebular $T_{\rm e}$ diagnostic line ratios from the ab initio 
photoionization models. Models using SEDs from \cmfgen, \tlusty, 
\wmbasic, and \fastwind\ are plotted as black, red, blue, 
and green curves, respectively. Vertical lines show the effect a variation of 
1000K in the electron temperature on the line ratios.}
\label{fig5}
\end{figure}
%
Figure \ref{fig5} illustrates the effect of the various SEDs on the 
nebular temperature from a different point of view. Here we plot the values  
of the \Te\  indicators \linean{S}{iii}{6312}/\linean{S}{iii}{9532} (left) and
\linean{N}{ii}{5755}/\linean{N}{ii}{6584} (right) as would be measured through 
a small square slit as a function of 
the projected distance to the central star, i. e. after integrating the line 
emissivities along a chord through the spherical
\hii\ region. Following the same set of
colors used along the paper, models using SEDs from \cmfgen, \tlusty, \wmbasic,
and \fastwind\ are represented in black, red, blue, and green 
lines, respectively. The electron temperatures that would be derived 
from a given \Te-indicator differ by about 200-300 K at most.
%
\subsection{Effect on the nebular ionization structure}
\label{section32}
%
Contrary to the energy gains, which  result from an integration over 
energy of the ionizing  photons, the nebular ionization structure depends on the 
\emph {details} of the SED. One can thus suspect that differences in the SEDs 
obtained with various stellar atmosphere codes will translate into significant 
differences in the ionization structures of the nebulae.
%
\subsubsection{General considerations}
\label{section321}
%
The state of ionization of a nebula depends only, in first approximation, on the 
ionization parameter, U, and on the hardness of the ionizing radiation. Following 
Vilchez \& Pagel (\citeyear{Vil88}),
\begin{equation}\label{equation4}
\frac{n({\rm X}\,^{i+1})}{n({\rm X}\,^{i})} \propto U 
     \frac{\int_{\nu(X^i)}^{\infty}\frac{H_\nu}{h\nu}d\nu}
	      {\int_{13.6 eV}^{\infty}\frac{H_\nu}{h\nu}d\nu} \propto U
		  \frac{{\rm Q(X}\,^{i})}{{\rm Q(H}\,^{0})},
\end{equation}
where $n$(\ion{X}{$^i$}) is the number density of the $i$-times ionized atoms of 
element X, and $H_{\nu}$ is the stellar Eddington flux. Therefore, a 
ratio $n$(\ion{X}{$^{i+1}$})/$n$(\ion{X}{$^i$}) is, to a first order and for a 
given U, proportional  to the relative number of photons able to ionize 
\ion{X}{$^i$}, as compared to that of Lyman continuum photons.
Note, however, that photoionization models show that 
Eq. \ref{equation4} is not exactly fulfilled (Stasi\'nska \& Schaerer 
\citeyear{Sta97}) rendering the interpretation more complex. Indeed, Eq. 
\ref{equation4} ignores the effect of the modification of the radiation field 
in the nebula due to nebular emission and absorption.
%
\subsubsection{Quantitative effects on the nebular ionization 
structure}
\label{section322}
%
%
%
\begin{table}
\begin{center}
\caption{Integrated ionic ratios (in logarithm) predicted 
by \cloudy\ photoionization models using the SEDs from the stellar models considered in 
Sect. \ref{section23}  
\label{t6}
}
\begin{tabular}{c c c c c c}    
\hline
\hline
 &  \ion{He}{$^{+}$}/\ \ion{H}{$^{+}$} &
    \ion{O}{$^{2+}$}/\ \ion{O}{$^{+}$} &
    \ion{Ne}{$^{2+}$}/\ \ion{Ne}{$^{+}$} &
    \ion{S}{$^{3+}$}/\ \ion{S}{$^{2+}$} &
    \ion{Ar}{$^{2+}$}/\ \ion{Ar}{$^{+}$} \\
\hline
C1 &   -2.163 &   -2.535 &   -3.599 &	-4.059 &   -1.249  \\
T1 &   -2.130 &   -2.663 &   -3.842 &	-4.235 &   -1.209  \\
W1 &   -2.087 &   -2.368 &   -3.263 &	-3.869 &   -1.166  \\
F1 &   -2.205 &   -2.561 &   -3.415 &	-4.040 &   -1.303  \\
\noalign{\smallskip}
C2 &   -1.293 &   -0.618 &   -1.775 &	-1.906 &   -0.020  \\
T2 &   -1.311 &   -0.885 &   -1.814 &	-2.291 &   -0.058  \\
W2 &   -1.531 &   -0.932 &   -1.777 &	-2.224 &   -0.418  \\
F2 &   -1.639 &   -1.346 &   -2.079 &	-2.796 &   -0.573  \\
\noalign{\smallskip}
C3 &   -1.012 &    0.531 &   -0.375 &   -0.777 &    1.528  \\
T3 &   -1.016 &    0.173 &   -0.587 &   -1.166 &    1.379  \\
W3 &   -1.012 &    0.506 &    0.144 &   -0.799 &    1.493  \\
F3 &   -1.047 &   -0.173 &   -0.838 &   -1.537 &    0.899  \\
\hline
\hline
\end{tabular}
\\
\end{center}
\end{table}
%
%
%
\begin{table}
\begin{center}
\caption{Results for the \ion{N}{$^{+}$}/\ \ion{O}{$^{+}$}
and \ion{Ne}{$^{2+}$}/\ \ion{O}{$^{2+}$} ionic ratios predicted by 
\cloudy\ photoionization models using the SEDs from 
the stellar models considered in Sect. \ref{section23}  
\label{t7}
}
\begin{tabular}{c c c}    
\hline
\hline
  & (\ion{N}{$^{+}$}/\ \ion{O}{$^{+}$}) / (N/ O) &
    (\ion{Ne}{$^{2+}$}/\ \ion{O}{$^{2+}$}) / (Ne/ O) \\
\hline
C1 &   1.002 &  0.086 \\
T1 &   1.000 &  0.066 \\
W1 &   0.998 &  0.127 \\
F1 &   1.002 &  0.139 \\
\noalign{\smallskip}
C2 &   0.804 &  0.085 \\
T2 &   0.748 &  0.132 \\
W2 &   0.899 &  0.157 \\
F2 &   0.910 &  0.192 \\
\noalign{\smallskip}
C3 &   0.668 &  0.385 \\
T3 &   0.445 &  0.345 \\
W3 &   0.612 &  0.767 \\
F3 &   0.422 &  0.317 \\
\hline
\hline
\end{tabular}
\\
\end{center}
\end{table}
%
Table \ref{t6} summarizes the predictions resulting from \cloudy\ models
using the stellar SED from the various stellar atmosphere codes in
terms of the ionic fractions (in logarithm) for several elements. A comparison
of these quantities can give us an idea of the effect that differences
in the stellar SEDs are producing on the nebular ionization structure.
As expected, in some cases photoionization models are resulting in enormous 
differences in these values for these quantities (up to $\sim$\,1 dex
for some cases). Obviously, there is a direct relation between how these 
quantities compare and the discrepancies found in Fig. \ref{fig1} for the 
stellar ionizing SEDs. For example, the largest difference in the ratio
\qion{He}{0} / \qion{H}{0} were found between \fastwind\ and \cmfgen\
for the 35000 K star case ($\sim$\,0.4 dex) and, consequently,
the largest difference in the ratio \ion{He}{$^+$}/\ion{H}{$^+$} 
predicted by the photoionization models is also found for the same
case (0.35 dex difference). 

The fact that different SEDs produce  different ionization structures has  
important consequences on abundance determinations and on the evaluation 
of the mean effective temperature of the ionizing stars. 

\emph {Effects on ionic ratios.} In Table \ref{t7} we consider the nebular ratios 
\ion{N}{$^+$}/\,\ion{O}{$^+$} and \ion{Ne}{$^{2+}$}/\,\ion{O}{$^{2+}$}. 
 A common way to derive N abundances
in \hii\ regions is to use [\ion{N}{ii}] and 
[\ion{O}{ii}] lines, and correct for the unseen \ion{N}{$^{2+}$} assuming that:
N/\,O\,=\,\ion{N}{$^+$}/\,\ion{O}{$^+$}. Similarly, the Ne abundance is obtained 
assuming that 
Ne/\,O\,=\,\ion{Ne}{$^{2+}$}/\,\ion{O}{$^{2+}$}. Both formulae are recipes based 
on similarities of ionization potentials.
Table \ref{t7} shows that not only these formulae are not necessarily correct, but 
also that the value of these ratios depend on the considered SEDs, even the ionization 
potentials of \ion{N}{$^+$} and \ion{O}{$^+$} are so close to each other (as seen in Fig. 
\ref{fig1}).

\begin{figure*}
\centering
\includegraphics[width=9.5 cm,angle=90]{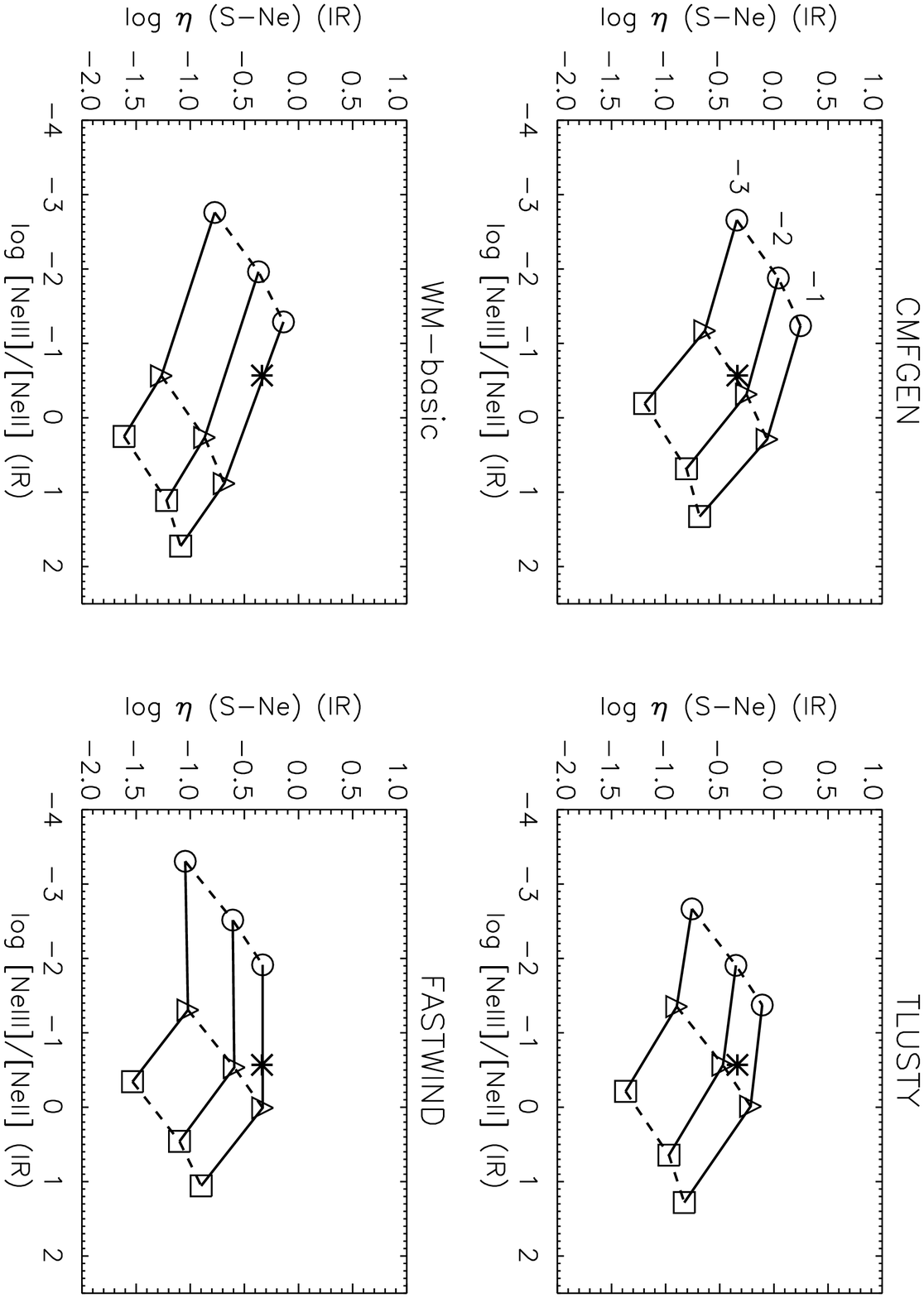}
\caption{Diagnostic diagrams proposed by Morisset (\citeyear{Mor04}) to derive 
$<$\Teff$>$ and U by comparing nebular infrared data with results from 
grids of photoionization models. The corresponding IR lines used for the
diagrams are [\ion{S}{iv}]\,10.5$\mu$m, [\ion{S}{iii}]\,18.7$\mu$m, 
[\ion{Ne}{iii}]\,15.5$\mu$m, and [\ion{Ne}{ii}]\,12.8$\mu$m.
The figure shows the position of the 
compact \hii\ region G29.96-0.02 studied by Morisset et al. (\citeyear{Mor02}) 
with respect to small grids of models constructed with each of the four 
stellar atmosphere codes under study. The grids were constructed using 
photoionization models with log\,$U$\,=\,-3,-2,-1 using fluxes from stellar
models with\ Teff\,=\,35, 40, and 45 kK (circles, triangles and squares, 
respectively).}
\label{fig6}
\end{figure*}

\emph {Estimation of the mean temperature of the ionizing stars.} This can be done 
by considering line ratios arising from different ions of the same element. 
The most popular indicator is the radiation softness parameter 
$\eta$\,=\,(\ion{O}{$^+$} /\,\ion{O}{$^{2+}$})\,/\,(\ion{S}{$^+$} /\,\ion{S}{$^{2+}$}) 
proposed by Vilchez \& Pagel (1988). A better way to estimate the hardness 
of the ionizing radiation field is to plot the observational points in the  
$\eta$(S-Ne)=([\ion{S}{iv}]/[\ion{S}{iii}])/([\ion{Ne}{iii}]/[\ion{Ne}{ii}]) 
vs. [\ion{Ne}{iii}]/[\ion{Ne}{ii}] plane, as shown by  Morisset (\citeyear{Mor04b}).  
$<$\Teff$>$ is then derived, 
together with  the ionization parameter $U$, by comparison with grids of 
models. In Fig. \ref{fig6}, we show the position of the compact \hii\ 
region G29.96-0.02 (Morisset et al. \citeyear{Mor02}) with respect to small 
grids of models constructed with each of the four stellar atmosphere 
codes under study. In this case we have excluded the 30000 K star case
and consider additional stellar models\footnote{These models were computed following the
same ideas as those presented in Sect. \ref{section21}.} with \Teff\,=45000 K 
and \grav\,=\,4.0 dex. 
We see that, while \cmfgen\ and \tlusty\  would infer  
$<$\Teff$>$ $\simeq$ 40000K, \wmbasic\ would imply $<$\Teff$>$ $\sim$ 37000K, 
and \fastwind\ would give $<$\Teff$>$ $\sim$ 38500K 

The differences between the values of  $<$\Teff$>$  obtained using different codes are   
due not only to the difference in the general slopes of the SEDs, but also to local 
properties of the SEDs close to the ionization potentials of the involved ions. 

If diagrams like those of Fig. \ref{fig6} are used to simply order the values of 
$<$\Teff$>$ of a sample of  \hii\ regions, one will probably obtain roughly the same 
ordering whatever atmosphere code is used (provided that one uses a diagram of 
appropriate metallicity for each object$<$\Teff$>$, as described by Morisset 
\citeyear{Mor04b}). But, as Fig. \ref{fig6} shows, taking as granted the absolute values 
of 
$<$\Teff$>$ is risky, since it depends so much on the stellar atmospheres used to 
determine them.

An additional problem is that the results depend strongly on whether the diagrams are 
constructed using models of dwarf stars or of supergiants. In general, for a given \Teff\ 
the ionizing SED is
harder in a star with lower gravity; therefore, lower stellar temperatures
will be derived if based on models of supergiants. This is the reason why we obtained a 
larger $<$\Teff$>$ for the ionizing star of  
G29.96-0.02 than Morisset et al. (\citeyear{Mor02}), since they used
stellar models with lower gravities. 

%
%
%

%
\section{Summary}
\label{section4}
%
The aim of this paper was to study how various modern 
stellar atmosphere codes compare in terms of ionizing SEDs and how this affects 
the emission line spectra of surrounding nebulae. This is  
important, since different codes predict quite different ionizing SEDs, 
which, in turn, impinges on the quantitative interpretation of \hii\ regions 
spectra in terms  of formation history, temperature of ionizing radiation  
or chemical composition. 

We have computed models using four state-of-the-art
stellar atmosphere codes: 
\tlusty, \cmfgen, \wmbasic, and \fastwind. All these codes account for  non-LTE
and line blanketing effects, but with different emphasis on specific aspects. 
\tlusty\ is a plane parallel, hydrostatic code, while the other three belong
to the family of models atmosphere codes so-called "unified models". On the 
other hand, while \tlusty\ and \cmfgen\ consider a fully consistent treatment of 
line blanketing, \wmbasic\ and \fastwind\ use a fast numerical method to account 
for the line blanketing and blocking effects.
Three sets of stellar parameters representing a late O-type (O9.5\,V), a mid 
O-type (O7\,V), and an early O-type (O5.5\,V) dwarfs were considered, with the 
chemical composition and microturbulence being exactly the same in all models, 
so as to perfectly isolate the effects on physics and numerical prescriptions 
from the effect of secondary input parameters.

We have shown that, in order to compare the ionizing SEDs predicted by 
various stellar atmosphere codes, it is important to distinguish between what we called 
\emph{evolutionary} and \emph{observational} approaches. The 
\emph{evolutionary} approach 
compares SEDs of stars defined by the same values of \Teff\ and \grav, and is 
relevant for studies based on stellar population synthesis. On the other hand, it is the 
\emph{observational} approach which is needed when using \hii\ regions to probe 
theoretical SEDs. In this approach, the models to be compared do not 
necessarily have the same \Teff\ and \grav, but they produce similar intensities and 
profiles for the stellar H and \ion{He}{i-ii} lines normally used to estimate the 
stellar parameters. Note that a combined tailored modelling of a massive star and 
its surrounding ionized nebula would precisely use the \emph{observational} approach. 

We have shown that the study of the optical H and \ion{He}{i-ii} stellar lines 
helps understanding the origin of differences in the ionizing SEDs predicted 
by various stellar atmosphere codes. There is a better agreement -- in terms 
of the total number of H ionizing photons \qion{H}{0}, of the ratio 
\qion{He}{0}/\qion{H}{0}, and of the shape of the SED in the spectral range 
13\,-\,30 eV --
between models resulting in similar H and \ion{He}{i-ii} lines than between 
models constructed with exactly the same \Teff\ and \grav.  
Therefore, to probe the ionizing radiation from massive stars by using 
nebular lines from surrounding \hii\ regions\footnote{Table \ref{tB1} in Appendix 
\ref{app2} (accessible on-line only) indicates the available optical 
and IR nebular line ratios useful to test the predicted emergent 
ionizing SEDs.}, it  is important to secure 
high quality spectra of the ionizing stars,  allowing one to accurately 
determine the stellar parameters. Photoionization modeling will then permit 
one to investigate the ionizing SED above 30 eV (where the shape of the SED 
is known to be very sensitive to the treatment of 
line blocking, line blanketing and stellar wind effects). \\ 

We then examined the emission line spectra of ab initio models of  \hii\ 
regions constructed with the various atmosphere codes. 
We found that the 
nebular temperatures do not differ by more than 300-350 K, and that the 
changes are not always in the direction expected from the change in the 
heating rate: differences in the cooling rate, induced by  the modification 
of the ionization structure, also play a role. The ionization structure 
itself may change substantially. This has an impact on the abundance 
ratios derived from \hii\ regions spectra, either by tailored photoionization 
modelling or with the help of ionization correction factors. For example, 
we have found that the
values of \ion{N}{$^{+}$}/\ \ion{O}{$^{+}$}
and \ion{Ne}{$^{2+}$}/\ \ion{O}{$^{2+}$}, commonly identified with the N/O and
Ne/O abundance ratios, may differ by 0.2--0.3 dex from one SED to another.
The changes in the ionization structure also affects the temperature of 
the ionizing stars estimated from the emission line ratios produced by the 
nebula. In the example we considered, the estimated \Teff\ varies from 
$\simeq$ 37000K to 40000K. \\

Note that, in this paper, we presented  models for dwarf stars only. For supergiant 
stars, the comparative behaviour of models obtained with the various codes  may be 
different.  For example, while Morisset et al. (\citeyear{Mor04})
found that the difference between \wmbasic\ and \cmfgen\
models for supergiants with \Teff\,=\,40000\,K is positive in the 
$>$35 eV range,  we found the  opposite for dwarfs with the same
\Teff\ (see also Fig. 8 in Morisset et al. \citeyear{Mor04}).
It is likely that the behavior will change also with the chemical composition. 

The problem is therefore complex and has to be analyzed further.  In any case, the 
limited study presented in this paper shows that, than even with the latest generation of
stellar atmosphere codes, one must be careful when combining them with 
photoionization codes to establish the ionization correction factors to be
used in the abundance determination of \hii\ regions, or to characterize 
the ionizing stellar population from nebular line ratios.

The follow-up of this study will be to analyze in detail a few Galactic \hii\ 
regions and their central stars, performing a combined stellar and nebular 
modelling, in order  check which of the stellar atmosphere codes provides 
the best ionizing SED. 
Hopefully, the present study and our forthcoming papers will have an impact 
on our understanding of the physics of stellar atmospheres, as well as on 
the reliability of the predictions from stellar synthesis codes and of 
abundance determinations in \hii\ regions. 
%

\section*{Acknowledgments}

We are extremely grateful to A. Herrero, F. Najarro, L. Mart\'in-H\'ernandez, 
and J. Garc\'ia-Rojas for their careful reading of the first submitted draft 
of the paper. Special thanks to J. Puls and V. Luridiana. We also thank to 
F.N., F. Martins, and T. Lanz for computing for us some \cmfgen\ and \tlusty\ 
models. We would also like to thank the referee, C. Morisset, for valuable
comments and suggestions to improve the final version of the paper.
This work has been funded by the Spanish Ministerio de Educaci\'on y 
Ciencia under the MEC/Fullbright postdoctoral fellowship program.

\appendix

\section{General characteristics of the stellar atmosphere codes}
\label{app1}
%
Below we summarize the main characteristics of the four stellar atmosphere 
codes considered in this study (we also refer to Table 1 in 
Herrero \citeyear{Her07} and Puls \citeyear{Pul08}) and give some notes on how the 
specific calculations of the models used here were performed.

Generally speaking, the four codes account for the \nlte\ and line 
blanketing effects (although they also offer the possibility of calculating 
\lte\ and/or unblanketed models). \tlusty\ is a plane parallel, 
hydrostatic code, while the other three (\fastwind, \wmbasic, and 
\cmfgen) belong to the family of model atmosphere codes so-called 
``unified models" (Gabler et al. \citeyear{Gab89}), in which a spherically 
symmetric geometry, with a smooth transition between the photosphere 
and the wind is considered.
%
\subsection{TLUSTY}
\label{app11}
%
\tlusty\ is a plane parallel hydrostatic code widely used nowadays.
This code is aimed at the study of spectral lines formed in the
photosphere, where velocities are small and the geometrical extension
is negligible. The model atmosphere code is completed with \synspect, 
a program for calculating the spectrum emerging from a given model atmosphere.

The basic concepts, equations and numerical methods used are described 
in Hubeny (\citeyear{Hub88}). New developments included in subsequent 
versions are described in detail in Hubeny \& Lanz (\citeyear{Hub92}, 
\citeyear{Hub95}) and Hubeny et al. (\citeyear{Hub94}). One of the most 
important features of the new version of the program (v. 200) is that 
it allows for a fully consistent, \nlte\ treatment of metal line 
blanketing, either by means of Opacity Distribution Functions (ODF) 
or Opacity Sampling (OS).

The program is fully data oriented as far as the choice of atomic 
species, ions, energy levels, transitions and opacity sources is 
concerned (there are no default opacities built in). Note that 
to calculate realistic line blanketed models with \tlusty\ 
all the above mentioned atomic data need to be included explicitly
in the calculations. This makes the model calculation to take
about 4 hours.
\newline
In order to construct realistic \nlte\ model atmospheres for O-type 
stars, Lanz \& Hubeny (\citeyear{Lan03}) included about 100000 individual 
atomic levels of over 40 ions of H, He, C, N, O, Ne, Si, P, S, Fe and 
Ni in the calculation of their {\sc Ostar2002} grid. The levels are 
grouped into about 900 \nlte\ superlevels. A total of 8000 lines of 
the light elements and about 2 million lines of 
\ion{Fe}{iii}-\ion{Fe}{vi} and \ion{Ni}{iii}-\ion{Ni}{vi} are accounted 
for in the calculations.

As commented in Sect. \ref{section21}, we used three models from the 
{\sc Ostar2002} grid for our study.

\tlusty\ and \synspect\ codes, the {\sc Ostar2002} grid and all the 
atomic data files used in it can be downloaded from 
{\tt http://nova.astro.umd.edu/}. There is also a very complete 
user's guide available. 
%
\subsection{CMFGEN}
\label{app12}
%
\cmfgen\ was originally 
designed for the analysis of W-R stars and LBVs, with very dense winds. 
Subsequent developments of the code have allowed to also extend its 
application to O stars. Nowadays, \cmfgen\ can be used for detailed 
spectroscopic studies of hot stars from the {\sc (e)uv} to the {\sc (f)ir} 
ranges.

The approach driving this code is the inclusion of as many lines as 
possible in order to fully describe the effects of line blanketing, 
while minimizing the number of level populations to be explicitly solved 
(using the idea of superlevels, first pioneered by Anderson \citeyear{And89}).
Radiative transfer is treated ``exactly" in this code (all lines are 
treated in comoving frame, even those from iron group elements), meaning 
that no opacity redistribution or sampling techniques are used.
Consequently, \cmfgen\ is very time consuming (model calculation takes 
anything from 3-12 hours, depending on the star and how close the 
initial guess is from the final solution).

Since \cmfgen\ was originally designed for the analysis of stars with 
very dense winds, the treatment of the photospheric density 
stratification is considered in an approximated way. At present, the 
velocity and density structures are not self-consistently computed but 
have to be given as input, either as input data or parameterized (see 
in Bouret et al. \citeyear{Bou03}, an example of how these structures are 
included in the code).

The \cmfgen\ models used in this study were kindly generated by F. Najarro 
and F. Martins. 
They used as initial guess for the model calculation previously calculated
models with stellar parameters close to the required ones (see Table
\ref{t1}). Each model calculation took approximately 1 day.
%
\subsection{WM-basic}
\label{app13}
%
\wmbasic\ aims mainly at the prediction of {\sc euv/uv} fluxes and 
profiles for hot luminous stars with homogeneous, stationary, and 
spherically symmetric radiation driven winds. In terms of the emergent 
spectrum calculation, in \wmbasic\ models the bound-bound radiative 
rates are calculated in Sobolev approximation and the code does not 
include Stark broadening. Therefore, 
in \wmbasic, the optical lines usually considered for the stellar 
parameter determination are not reliable for diagnostic purposes.

\wmbasic\ provides a realistic stratification of the density and 
velocity, particularly in the transonic region, through the solution 
of the hydrodynamics produced by the scattering and absorption of 
Doppler shifted metal lines. The line contribution to the radiative 
acceleration is parameterized following the formulation by Castor, 
Abbot \& Klein (\citeyear{Cas75}) --- CAK ---  (see also Pauldrach et al. 
\citeyear{Pau86}) that uses the concept of line-force multiplier and the 
force multiplier parameters (K, $\alpha$ and $\delta$).

The authors have developed a fast numerical method which accounts 
for the blocking and blanketing influence of all metal lines in the 
entire sub- and supersonically expanding atmosphere in a consistent 
manner (see Pauldrach et al. \citeyear{Pau01}). This method makes 
the computational time to be significantly reduced in comparation with
e.g. \cmfgen\ calculations. To this end, detailed atomic models are 
implicitly considered by the code for all the important ions. These 
include 149 ionizations stages of 
26 elements (H to Zn, apart from Li, Be, B and Sc), resulting in 
a total of 5000 levels. More than 30000 bound-bound transitions for 
the \nlte\ calculations and more than 4 million lines for the 
line-force and blocking calculations are accounted for.


Our \wmbasic\ models were generated with the easy-to-use Windows
interface created by Adi Pauldrach for running the code\footnote{This
is distributed as freeware and can be downloaded from the website:
{\tt www.usm.uni-muenchen.de/people/adi/adi.html.}} The initial
guess of the force multiplier parameters (K, $\alpha$ and 
$\delta$) provided by \wmbasic\ was slightly modified to produce
the same $\dot{M}$ and $v_{\infty}$ values than the ones used for
\fastwind\ and \cmfgen\ models (see Table \ref{t1}). The models
took $\sim$\,30-50 min. to be converged.
%
\subsection{FASTWIND}
\label{app14}
%
\fastwind\ (Santolaya-Rey et al. \citeyear{San97}; Puls et al. \citeyear{Pul05}) 
is optimized for the analysis of optical and IR spectra of OBA-type stars 
of all luminosity classes and wind strengths. 

Among the main objectives that have guided the developers of \fastwind\ 
since the  initial version of the code a fast performance is the
highest priority motivation.
Since the first version, presented in Santolaya-Rey et al. (\citeyear{San97}) 
to the latest one (last described by Puls et al. \citeyear{Pul05}), many 
improvements have been included in the code (the main ones are an 
approximate treatment of the line blanketing/blocking effects and a 
consistent calculation of the temperature structure by means of the thermal 
balance of electrons). 

The required computational efficiency is obtained by applying 
appropriate physical approximations to processes where high accuracy 
is not required (regarding the objective of the analysis: optical/IR 
lines). One of these approximations concerns the treatment of the 
opacity from metals (line blocking). We refer to Puls et al. 
(\citeyear{Pul05}) for a detailed description of the method. 

The code distinguishes between explicit elements, that are included 
in detail and can be synthesized later when solving the formal solution,
and background elements producing the line blanketing and whose occupation
numbers are calculated by solving the \nlte\ bound-bound rate equations in
Sobolev approximation, in a similar way as \wmbasic\ does.
The calculation of the line blanketing effects is
hence independent of the elements considered explicitly. 

Data from the background ions are taken from Pauldrach
et al. (\citeyear{Pau01}; \wmbasic) and are provided in a fixed form.\\

Although \fastwind\ is not aimed at predicting accurate ionizing spectral 
energy distributions (due to the simplified treatment of line blocking), 
we included it in our comparisons to test the reliability of the approximate 
SED predicted by this code.
\fastwind\ models were calculated with version v8.5 of the code. 
Since we were only interested in the SED and the optical H and He
lines, we considered H and He as the only explicit elements.
This way we could generate \fastwind\ models in only $\sim$\,10-20 min.
%
\section{Nebular line ratios better suited to test the shape of 
the stellar ionizing SED}
\label{app2}
%
%
%
\begin{table*}
\centering
\caption{Summary of nebular line ratios useful to test the 
         predicted emergent ionizing SEDs. 
\label{tB1}
}
\scriptsize{
\begin{tabular}{c c c c c c c c}    
\hline
\hline
Ioniz. edge & E$_{\rm ioniz}$ (eV) & $\lambda$ (\AA) & Ionic ratio & Line ratio & Optical 
range (\AA) & IR range ($\mu$m) & Other lines \\
\hline
\ion{S}{$^+$}     & 23.3 & 531 & \xion{S}{2+}{+}   & \xnion{S}{iii}{ii}  &  
\xlinean{S}{iii}{9531}{S}{ii}{6716+30} & ---  & \linean{S}{iii}{9069, 6312} \\
\ion{Cl}{$^+$}    & 23.8 & 520 & \xion{Cl}{2+}{+}  & \xnion{Cl}{iii}{ii} &   
\xlinean{Cl}{iii}{5517+37}{Cl}{ii}{8579}   & ---  & \linean{Cl}{ii}{9123, 6161} \\
\hline
\ion{He}{$^0$}    & 24.6 & 504 & \ion{He}{$^+$}\,/ \ion{H}{$^+$} & \ion{He}{i}\,/ 
\ion{H}{i} & \linea{He}{i}{5876}\,/ H$\beta$\,\footnote{There are many other possible 
combinations of \ion{He}{i} and \ion{H}{i} lines
available, both in the optical and the IR.}
 & & \\
\hline
\ion{Ar}{$^+$}    & 27.6 & 448 & \xion{Ar}{2+}{+}  & \xnion{Ar}{iii}{ii} & --- & 
\xlinean{Ar}{iii}{9.0}{Ar}{ii}{7.0}     & \linean{Ar}{iii}{21.8} \\
\ion{N}{$^+$}     & 29.6 & 419 & \xion{N}{2+}{+}   & \xnion{N}{iii}{ii}  & --- & 
\xlinean{N}{iii}{57.3}{N}{ii}{121.7}          &                             \\
\ion{S}{$^{2+}$}  & 34.8 & 356 & \xion{S}{3+}{2+}  & \xnion{S}{iv}{iii}  & --- & 
\xlinean{S}{iv}{10.5}{S}{iii}{18.7}     & \linean{S}{iii}{33.5}  \\
\ion{O}{$^+$}     & 35.1 & 353 & \xion{O}{2+}{+}   & \xnion{O}{iii}{ii}  & 
\xlinean{O}{iii}{5007}{O}{ii}{3726+29} & ---  & \linean{O}{iii}{4959, 4363}, 
\linean{O}{ii}{7320+30}\\
\ion{Cl}{$^{2+}$} & 39.6 & 312 & \xion{Cl}{3+}{2+} & \xnion{Cl}{iv}{iii} & 
\xlinean{Cl}{iv}{8048}{Cl}{iii}{5517+37} & ---      &  \\
\ion{Ne}{$^+$}    & 40.5 & 304 & \xion{Ne}{2+}{+}  & \xnion{Ne}{iii}{ii} &  --- & 
\xlinean{Ne}{iii}{15.6}{Ne}{ii}{12.8}  & \linean{Ne}{iii}{36.0} \\
\ion{Ar}{$^{2+}$} & 40.7 & 302 & \xion{Ar}{3+}{2+} & \xnion{Ar}{iv}{iii} & 
\xlinean{Ar}{iv}{4711+40}{Ar}{iii}{7751} & ---      & \linean{Ar}{iv}{7170}, 
\linean{Ar}{iii}{7135} \\
\ion{S}{$^{3+}$}  & 47.3 & 262 & \xion{S}{4+}{3+}  & \xnion{S}{v}{iv}    &  --- & --- \\
\ion{N}{$^{2+}$}  & 47.5 & 261 & \xion{N}{3+}{2+}  & \xnion{N}{iv}{iii}  &  --- & --- \\
\hline
\hline
\end{tabular}
\\
}
\end{table*}
%
%
\hii\ regions emit numerous H and He recombination lines in the optical, 
along with some forbidden lines from abundant metals. In addition, a 
rich spectrum of H recombination lines and fine structure lines is 
observed from the near- to far- IR (Ne, S, Ar: $\sim$3-40 $\mu$m;  C, N, 
O $\ge$40 $\mu$m). We refer to Stasi\'nska (\citeyear{Sta07}), who gives a list 
of useful  lines from C, N, O, Ne, S, Cl, and Ar ions and shows how to deal 
with them. Using appropriate nebular line ratios and taking into account the 
ideas presented above one can impose constraints to test the predicted 
emergent ionizing SEDs. \\

Nebular line ratios involving a \ion{He}{i} recombination line relative 
to a \ion{H}{i} line (e.g. \linea{He}{i}{5876}\,/ H$\beta$, in the 
optical) are important constraints. 
In an ionization bounded nebula, such a line ratio depends almost 
exclusively on the ratio \qion{He}{0}\,/\,\qion{H}{0}, provided that 
helium is not ionized in the entire nebula (which occurs if the effective
temperature of the stellar ionizing source is $\ge$\,40000 K).

Intensity ratios of lines from successive ions (\ion{X}{$^{i+1}$} and 
\ion{X}{$^i$}) of the same element can be used to constrain the ratio 
of ionic abundances and, as commented in Sect. \ref{section321}, the
number of stellar photons able to ionize \ion{X}{$^i$} relative to the
number of Lyman photons, \qion{X}{i}/\qion{H}{0}. In order to  isolate 
the effect of the SEDs on the ionization structure from the effect on 
nebular temperature, it is better to use pairs of lines which have a 
similar dependence on the electron temperature, i.e. either pairs of 
recombination lines, or pairs of far IR lines, or pairs of lines with 
similar excitation potentials (e. g. 
\linean{S}{iii}{9069}/\linean{S}{ii}{6716+30} rather than 
\linean{S}{iii}{6312}/\linean{S}{ii}{6716+30}). 

Table \ref{tB1} gives the list of available nebular line ratios.
For N, O, Ne, S, Cl, and Ar, only nebular and fine structure lines 
resulting from the lower energy levels are considered. 
In those cases when various lines are available in the optical or
IR ranges, the most intense ones are selected, though other 
possible lines are also indicated in the last column. 
%
%
%

\bsp

\label{lastpage}


\begin{thebibliography}{}
\bibitem[\protect\citeauthoryear{}{1989}]{And89}
Anderson L.~S., 1989, ApJ, 339, 558 
\bibitem[\protect\citeauthoryear{}{2003}]{Bou03}
Bouret, J.-C., Lanz, T., Hillier, D.~J., Heap, S.~R., Hubeny, I., Lennon, D.~J., 
Smith, L.~J., \& Evans, C.~J.\ 2003, ApJ, 595, 1182 
\bibitem[\protect\citeauthoryear{}{1975}]{Cas75}
Castor, J.~I., Abbott, D.~C., \& Klein, R.~I.\ 1975, ApJ, 195, 157
\bibitem[\protect\citeauthoryear{}{2004}]{Est04}
Esteban, C., Peimbert, M., Garc{\'{\i}}a-Rojas, J., Ruiz, M.~T., Peimbert, A., \& 
Rodr{\'{\i}}guez, M.\ 2004, MNRAS, 355, 229 
\bibitem[\protect\citeauthoryear{}{1998}]{Fer98}
Ferland, G.~J., Korista, K.~T., Verner, D.~A., Ferguson, J.~W., Kingdon, J.~B., \& 
Verner, E.~M.\ 1998, PASP, 110, 761 
\bibitem[\protect\citeauthoryear{}{1999}]{Fio99}
Fioc, M., \& Rocca-Volmerange, B.\ 1999, ArXiv Astrophysics e-prints, 
arXiv:astro-ph/9912179 
\bibitem[\protect\citeauthoryear{}{1989}]{Gab89}
Gabler, R., Gabler, A., Kudritzki, R.~P., Puls, J., \& Pauldrach, A.\ 1989, A\&A, 226, 
162
\bibitem[\protect\citeauthoryear{}{2002}]{Giv02}
Giveon, U., Sternberg, A., Lutz, D., Feuchtgruber, H., \& Pauldrach, A.~W.~A.\ 2002, ApJ, 
566, 880 
\bibitem[\protect\citeauthoryear{}{1998}]{Gre98}
Grevesse, N., \& Sauval, A.~J.\ 1998, Space Science Reviews, 85, 161 
\bibitem[\protect\citeauthoryear{}{1992}]{Her92}
Herrero, A., Kudritzki, R.~P., Vilchez, J.~M., Kunze, D., Butler, K., \& Haser, S.\ 1992, 
A\&A, 261, 209 
\bibitem[\protect\citeauthoryear{}{2002}]{Her02}
Herrero, A., Puls, J., \& Najarro, F. 2002, A\&A, 396, 949
\bibitem[\protect\citeauthoryear{}{2007}]{Her07}
Herrero, A.\ 2007, ArXiv e-prints, 704, arXiv:0704.3528 
\bibitem[\protect\citeauthoryear{}{1998}]{Hil98}
Hillier, D.~J., \& Miller, D.~L.\ 1998, ApJ, 496, 407 
\bibitem[\protect\citeauthoryear{}{1988}]{Hub88}
Hubeny, I.\ 1988, Computer Physics Comm. 52, 103
\bibitem[\protect\citeauthoryear{}{1992}]{Hub92}
Hubeny, I., \& Lanz, T.\ 1992, A\&A, 262, 501 
\bibitem[\protect\citeauthoryear{}{1994}]{Hub94}
Hubeny, I., Hummer, D.~G., \& Lanz, T.\ 1994, A\&A, 282, 151 
\bibitem[\protect\citeauthoryear{Hubeny 
\& Lanz}{1995}]{Hub95} Hubeny I., Lanz T., 1995, ApJ, 439, 875 
\bibitem[\protect\citeauthoryear{}{2005}]{Jam05}
Jamet, L., Stasi{\'n}ska, G., P{\'e}rez, E., Gonz{\'a}lez Delgado, R.~M., \& 
V{\'{\i}}lchez, J.~M.\ 2005, A\&A, 444, 723 
\bibitem[\protect\citeauthoryear{}{1980}]{Kud80} 
Kudritzki, R.-P.\ 1980, A\&A, 85, 174 
\bibitem[\protect\citeauthoryear{}{2000}]{Kud00} 
Kudritzki R.-P., Puls J., 2000, ARA\&A, 38, 613 
\bibitem[\protect\citeauthoryear{}{1992}]{Kun92}
Kunze, D., Kudritzki, 
R.-P., \& Puls, J.\ 1992, The Atmospheres of Early-Type Stars, 401, 45
\bibitem[\protect\citeauthoryear{}{1994}]{Kun94}
Kunze, D.\ 1994, Ph.D.~Thesis
\bibitem[\protect\citeauthoryear{}{1991}]{Kur91}
Kurucz, R.~L.\ 1991, BAAS, 23, 1047 
\bibitem[\protect\citeauthoryear{}{1992}]{Kur92}
Kurucz, R.~L.\ 1992, RMxAA, vol.~23, 23, 45 
\bibitem[\protect\citeauthoryear{}{2003}]{Lan03}
Lanz, T., \& Hubeny, I.\ 2003, ApJs, 146, 417 
\bibitem[\protect\citeauthoryear{}{1999}]{Lei99}
Leitherer, C., et al.\ 1999, ApJs, 123, 3 
\bibitem[\protect\citeauthoryear{}{1963}]{Mat63}
Matthews, T.~A., \& Sandage, A.~R.\ 1963, ApJ, 138, 30 
\bibitem[\protect\citeauthoryear{}{2002}]{Mart02}
Mart\'in-Hern\'andez, N. L., Vermeij, R., Tielens, A. G. G. M., van der Hulst, J. M., \& 
Peeters, E., 2002, A\&A, 389, 286
\bibitem[\protect\citeauthoryear{}{2005}]{Mar05}
Martins, F., Schaerer, D., \& Hillier, D.~J.\ 2005, A\&A, 436, 1049 
\bibitem[\protect\citeauthoryear{}{2000}]{Mey00}
Meynet, G., \& Maeder, A. 2000, A\&A 361, 101
\bibitem[\protect\citeauthoryear{}{2005}]{Mey05}
Meynet, G., \& Maeder, A. 2005, A\&A 429, 581
\bibitem[\protect\citeauthoryear{}{1970}]{Mih70}
Mihalas, D., \& Auer, L.~H.\ 1970, ApJ, 160, 1161 
\bibitem[\protect\citeauthoryear{}{2004}]{Mok04}
Mokiem, M.~R., Mart{\'{\i}}n-Hern{\'a}ndez, N.~L., Lenorzer, A., de Koter, A., \& 
Tielens, A.~G.~G.~M.\ 2004, A\&A, 419, 319 
\bibitem[\protect\citeauthoryear{}{2002}]{Mor02}
Morisset, C., Schaerer, D., Mart{\'{\i}}n-Hern{\'a}ndez, N.~L., Peeters, E., Damour, F., 
Baluteau, J.-P., Cox, P., \& Roelfsema, P.\ 2002, A\&A, 386, 558
\bibitem[\protect\citeauthoryear{}{2004}]{Mor04}
Morisset, C., Schaerer, D., Bouret, J.,-C., \& Martins, F. 2004, A\&A, 415, 577
\bibitem[\protect\citeauthoryear{}{2004}]{Mor04b} 
Morisset C., 2004, ApJ, 601, 858 
\bibitem[\protect\citeauthoryear{}{2005}]{Mor05}
Morisset, C., Stasi{\'n}ska, G., \& Pe{\~n}a, M.\ 2005, MNRAS, 360, 499 
\bibitem[\protect\citeauthoryear{}{1996}]{Naj96}
Najarro, F., Kudritzki, R.~P., Cassinelli, J.~P., Stahl, O., \& Hillier, D.~J.\ 1996, 
A\&A, 306, 892 
\bibitem[\protect\citeauthoryear{}{2006}]{Naj06}
Najarro, F., Hillier, D.,J., Puls, J., et al. 2006, A\&A, 456, 659
\bibitem[\protect\citeauthoryear{}{2000}]{Oey00}
Oey, M.~S., Dopita, M.~A., Shields, J.~C., \& Smith, R.~C.\ 2000, ApJs, 128, 511 
\bibitem[\protect\citeauthoryear{}{1986}]{Pau86}
Pauldrach, A., Puls, J., \& Kudritzki, R.~P.\ 1986, A\&A, 164, 86 
\bibitem[\protect\citeauthoryear{}{2001}]{Pau01}
Pauldrach, A.~W.~A., Hoffmann, T.~L., \& Lennon, M.\ 2001, A\&A, 375, 161 
\bibitem[\protect\citeauthoryear{}{2005}]{Pul05}
Puls J., Urbaneja M.,A., Venero R., et al. 2005, A\&A, 435, 669 
\bibitem[\protect\citeauthoryear{}{2008}]{Pul08}
Puls J., 2008, proceedings IAU250 "Massive Stars as Cosmic Enginees" 
\bibitem[\protect\citeauthoryear{}{2002}]{Rel02}
Rela{\~n}o, M., Peimbert, M., \& Beckman, J.\ 2002, ApJ, 564, 704 
\bibitem[\protect\citeauthoryear{}{2004}]{Rep04}
Repolust, T., Puls, J., \& Herrero, A. 2004, A\&A, 415, 349
\bibitem[\protect\citeauthoryear{}{2005}]{Rep05}
Repolust, T., Puls, J., Hanson, M.~M., Kudritzki, R.-P., \& Mokiem, M.~R.\ 2005, A\&A, 
440, 261 
\bibitem[\protect\citeauthoryear{}{1995}]{Rub95}
Rubin, R.~H., Kunze, D., \& Yamamoto, T.\ 1995, Astrophysical Applications of Powerful 
New Databases, 78, 479 
\bibitem[\protect\citeauthoryear{}{2007}]{Rub07}
Rubin, R.~H., et al.\ 2007, MNRAS, 377, 1407 
\bibitem[\protect\citeauthoryear{}{1997}]{San97}
Santolaya-Rey A.,E., Puls J., \& Herrero A. 1997, A\&A, 323, 488
\bibitem[\protect\citeauthoryear{}{2000}]{Sch00}
Schaerer, D.\ 2000, Stars, Gas and Dust in Galaxies: Exploring the Links, 221, 99 
\bibitem[\protect\citeauthoryear{}{1992}]{Sch92}
Schaller, G., Schaerer, D., Meynet, G., \& Maeder, A. 1992, A\&A S 96, 269
\bibitem[\protect\citeauthoryear{}{1996}]{Sel96}
Sellmaier, F.~H., Yamamoto, T., Pauldrach, A.~W.~A., \& Rubin, R.~H.\ 1996, A\&A, 305, 
L37 
\bibitem[\protect\citeauthoryear{}{2006}]{Sim06}
Sim{\'o}n-D{\'{\i}}az, S., Herrero, A., Esteban, C., \& Najarro, F.\ 2006, A\&A, 448, 351 
\bibitem[\protect\citeauthoryear{}{1997}]{Sta97}
Stasinska, G., \& Schaerer, D.\ 1997, A\&A, 322, 615 
\bibitem[\protect\citeauthoryear{}{2004}]{Sta04}
Stasi{\'n}ska, G.\ 2004, Cosmochemistry.~The melting pot of the elements, 115
\bibitem[\protect\citeauthoryear{}{2007}]{Sta07}
Stasinska, G.\ 2007, ArXiv e-prints, 704, arXiv:0704.0348 
\bibitem[\protect\citeauthoryear{}{2002}]{Smi02}
Smith, L.~J., Norris, R.~P.~F., \& Crowther, P.~A.\ 2002, MNRAS, 337, 1309
\bibitem[\protect\citeauthoryear{}{1996}]{Vac96}
Vacca, W.~D., Garmany, C.~D., \& Shull, J.~M.\ 1996, ApJ, 460, 914 
\bibitem[\protect\citeauthoryear{}{1988}]{Vil88}
Vilchez, J.~M., \& Pagel, B.~E.~J.\ 1988, MNRAS, 231, 257 
\bibitem[\protect\citeauthoryear{}{2000}]{Vil00}
Villamariz, M.~R., \& Herrero, A.\ 2000, A\&A, 357, 597 
\bibitem[\protect\citeauthoryear{}{1972}]{Wal72}
Walborn, N.~R.\ 1972, AJ, 77, 312 
%
\end{thebibliography}
\end{document}